\documentclass{article}
\usepackage{graphicx}
\usepackage{subcaption}
\usepackage{mathtools}
\usepackage{stmaryrd}
\usepackage{amssymb}
\usepackage{amsthm}
\usepackage{nccmath}
\usepackage{hyperref}
\usepackage{bbm}
\usepackage{appendix}

\usepackage{fullpage}
\usepackage{parskip}
\usepackage{xcolor}

\usepackage{natbib}
\usepackage{apalike}
\usepackage[para]{footmisc}
\usepackage{mathrsfs}

\usepackage{booktabs}
\usepackage{algorithm}
\usepackage{algpseudocode}

\title{Optimal risk-aware interest rates for decentralized lending protocols}
\author{Bastien Baude\footnote{\href{mailto:bastien.baude@centralesupelec.fr}{bastien.baude@centralesupelec.fr}} \quad Damien Challet\footnote{\href{mailto:damien.challet@centralesupelec.fr}{damien.challet@centralesupelec.fr}} \quad Ioane Muni Toke\footnote{\href{mailto:ioane.muni-toke@centralesupelec.fr}{ioane.muni-toke@centralesupelec.fr}} \\ \\ \textit{Université Paris-Saclay, CentraleSupélec, Laboratoire MICS} \\
\textit{91192 Gif-sur-Yvette, France}}

\begin{document}

\bibliographystyle{apalike}

\maketitle

\begin{abstract}
\setlength{\baselineskip}{10.5pt}
\noindent
Interest rates in decentralized lending protocols are set algorithmically and adjust to supply and demand for liquidity. In this study, we propose an optimal interest rate model that maximizes the expected lender wealth while incorporating penalties for liquidity risk and interest rate stabilization. This objective benefits both sides of the market: it improves yield and reduces liquidity risk for lenders, while encouraging borrower activity indirectly through higher utilization and directly through stabilized borrowing costs. The dynamics of the utilization rate are modeled using point processes whose intensities depend on the interest rate. When intensities are linear, the optimal interest rate model is derived from a system of Riccati-type ODEs. In the nonlinear case, we approximate it using a Monte-Carlo estimator coupled with deep learning techniques. Finally, using block-by-block data, we conduct a risk-adjusted profit and loss analysis to compare industry-standard interest rate models to the deep learning-based one.
\\

\noindent\textbf{Keywords} -- Decentralized finance, decentralized lending protocol, interest rates, liquidity risk, market microstructure, stochastic optimal control, deep learning.

\end{abstract}
\tableofcontents

\newpage

\section{Introduction}

Decentralized lending protocols are a cornerstone of the Decentralized Finance (DeFi) ecosystem \citep{capponi2023decentralized}. They enable lending and borrowing of crypto-assets without traditional intermediaries such as banks or insurance companies. Lenders contribute liquidity to the market to earn interest. Meanwhile, borrowers secure loans by providing crypto-assets as collateral. Loans are over-collateralized and subject to liquidation procedures to preserve protocol solvency.

The Total Value Locked (TVL) refers to the value of the crypto-assets, generally expressed in dollars, held in the protocol. For lending protocols, it includes both available liquidity and posted collateral. As of January 2026, the TVL across all lending protocols was $\$60 \text{b}$,\footnote{Data sourced from DefiLlama, available at \url{https://defillama.com/protocols/Lending}.} a tenfold increase since January 2021, reflecting significant growth and adoption.

The Aave \citep{whitepaper2020aavev1,whitepaper2020aavev2,frangella2022aave} protocol is a pioneer in the market and has introduced innovative features that have set industry standards. Launched in 2017, it accounted for $50 \%$ of the lending market’s total TVL as of January 2026. More recently, a new player has emerged in the market: Morpho \citep{whitepaper2023morpho}, which has experienced rapid growth, with a TVL reaching $10 \%$ of the market as of January 2026.

In contrast to traditional lending markets \citep{hull2014options}, decentralized protocols structure loans without a fixed maturity. Hence, the interest rate paid by borrowers fluctuates based on supply and demand for liquidity within the protocol. Most lending protocols use a simple formula to set the interest rate based on the utilization rate, defined as the ratio of borrowed to supplied funds. The kinked rate model, introduced in \citet{whitepaper2020aavev1}, is a widely adopted, non-decreasing piecewise-linear function of the utilization rate with two distinct slopes.

Few studies have investigated interest rate models and market dynamics in decentralized lending protocols. \citet{cohen2023economics} characterize the equilibrium between borrowers and lenders and propose an interest rate model aimed at driving the utilization rate toward a target level. They also discuss potential exploits where lenders can manipulate the utilization rate to increase their profits. Potential exploits and vulnerabilities are further examined in \citet{chitra2023attacks}. \citet{bastankhah2024thinking} propose a data-driven interest rate model that adapts to shifts in market behavior. \citet{bastankhah2024agilerate} additionally highlight that introducing adaptivity into interest rate models can enhance market efficiency, but may also increase its vulnerability to adversarial manipulation. An interest rate model is proposed in \citet{bertucci2024agents} as the optimal solution to a stochastic control problem with an infinite horizon. Their objective is to minimize the expected deviations from a target utilization rate, while preventing full utilization. The evolution of the utilization rate is driven by a Brownian motion, with a drift linearly dependent on the interest rate. The model is calibrated using daily data from Aave and Compound.

In the spirit of \citet{bertucci2024agents}, we introduce a model for a liquidity pool by proposing a dynamic of the utilization rate using point processes where the intensities depend on the interest rate. We then formulate a stochastic control problem whose optimal solution yields an interest rate model that maximizes expected lender wealth over a finite horizon, while incorporating risk penalties to account for liquidity risk and interest rate stabilization. Our choice of objective function benefits all participants in the protocol. Lenders benefit from higher yields and reduced liquidity risk. At the same time, maximizing lender wealth indirectly encourages borrower activity by promoting high utilization. In addition, the interest rate stabilization penalty prevents excessive borrowing costs.

When the intensities are linear functions of the interest rate, the optimal interest rate model is derived from a system of Riccati-type ODEs. For nonlinear intensity functions, we approximate it using a Monte-Carlo estimator coupled with deep learning techniques. The learning procedure is also adapted to parametric interest rate models commonly used in the industry: linear, kinked and adaptive \citep{adaptive2023morpho}. After calibrating each model using block-by-block historical data from the USDT liquidity pool of Aave over a short time window, we conduct a risk-adjusted Profit and Loss (PnL) analysis to compare these parametric models to the deep learning-based one.

Such stochastic control problems are typically solved numerically using finite difference schemes. However, these methods suffer from the so-called ``curse of dimensionality''. To address this challenge, deep neural network-based algorithms have been proposed in the literature. These methods aim to parametrize the feedback control, i.e., a function of the state process, using a deep neural network which is then trained on Monte-Carlo simulations \citep{han2016deep}; see \citet{hure2021deep, germain2021neural} for applications in finance. As highlighted in \citet{lu2024multi}, while the literature on deep learning algorithms for stochastic control problems under Brownian noise is extensive, stochastic control problems involving jumps have received only limited attention. We develop, to the best of our knowledge, the first deep neural network-based algorithm to approximate the optimal solution of a stochastic control problem with jump dynamics where intensities are control-dependent. Reinforcement learning algorithms \citep{sutton2018reinforcement} can be used as alternative methods for solving such stochastic control problems; see \citet{gueant2019deep, baldacci2019market} for applications in finance.

Three main findings emerge from the analysis. When the intensities are linear functions of the interest rate and for a very short maturity, the optimal interest rate model corresponds to the kinked one. For piecewise-linear intensities, the neural network-based interest rate model is also piecewise-linear, demonstrating its ability to adapt to diverse market conditions. In the considered market configuration, the adaptive rate model performs similarly to the deep learning-based one, while the kinked rate model lacks effective risk management, especially near full utilization, and the linear rate model performs the worst.

The remainder of the paper is organized as follows. Section \ref{section:section_optimal_interest_rate_models} formulates the risk-aware interest rate problem. We then derive closed-form optimal solutions in the linear case and propose a deep learning-based numerical algorithm in the nonlinear case. Section \ref{section:numerical_results} presents numerical results based on synthetic data and analyzes the impact of model parameters. Applications to market data from the USDT liquidity pool on Aave v3 are presented in Section \ref{section:applications} and we compare the performance of standard parametric interest rate models to the deep learning-based one.

\section{Optimal interest rate models for lending protocols}\label{section:section_optimal_interest_rate_models}

\subsection{Mathematical framework}\label{section:mathematical_framework}

The role of the lending protocol is to propose an interest rate in real-time, based on the state of the liquidity pool. The state of the liquidity pool is characterized by the total value supplied, denoted by $S_{t}$ and the total value borrowed, denoted by $B_{t}$. It follows that $0 < B_{t} \leq S_{t}$, as both quantities are non-negative and the borrowed amount cannot exceed the supplied amount. In this context, the utilization rate, denoted by $U_{t}$, is defined as the ratio of the borrowed funds to the supplied funds and is bounded between $0$ and $1$:
\begin{equation}\label{eq:utilization_rate}
U_{t} = \frac{B_{t}}{S_{t}}.
\end{equation}
The interest rate models currently employed by major lending platforms are functions on the utilization rate rather than the total values supplied and borrowed. This modeling choice is motivated by the objective of the lending protocol to remain as attractive as possible while maintaining sufficient liquidity to accommodate potential future withdrawals. Indeed, the protocol faces liquidity risk when available reserve is insufficient, which materializes when $U_{t}$ reaches $1$. In the spirit of \citet{bertucci2024agents}, we model the dynamics of the utilization rate, assumed to be given by:
\begin{equation}\label{eq:utilization_dynamics}
dU_{t} = \delta dN^{{+}}_{t} - \delta dN^{{-}}_{t},
\end{equation}
where $(N^{+}_{t})_{t \geq 0}$ and $(N^{-}_{t})_{t \geq 0}$ are two point processes characterized by the intensity processes $(\lambda^{+}_{t})_{t \geq 0}$ and $(\lambda^{-}_{t})_{t \geq 0}$, respectively. Variations in utilization are assumed to be of constant size, with $\delta > 0$. We work with a probability space $\big ( \Omega, \mathcal{F} = (\mathcal{F}_{t})_{t \geq 0}, \mathbb{P} \big )$ where $\mathcal{F}$ is the natural filtration generated by the point processes $(N^{+}_{t})_{t \geq 0}$ and $(N^{-}_{t})_{t \geq 0}$. The intensity processes depend on both the interest rate and the utilization rate:
\begin{equation}\label{eq:intensities}
\lambda^{+}_{t} = \lambda^{+}(r_{t}) \mathbbm{1}_{U_{t^{-}} < 1}, \quad \lambda^{-}_{t} = \lambda^{-}(r_{t}) \mathbbm{1}_{U_{t^{-}} > 0},
\end{equation}
where $(r_{t})_{t \geq 0} \in \mathcal{A}$ is the interest rate process and $\mathcal{A}$ the set of $\mathcal{F}$-predictable processes. This modeling choice ensures that the utilization rate remains bounded between $0$ and $1$. Furthermore, we assume that the function $\lambda^{+}$ (respectively $\lambda^{-}$) is non-increasing (respectively non-decreasing) in the interest rate $r_{t}$. Indeed, from a financial perspective, an increase in the interest rate results in a decrease in borrowing demand (and correspondingly an increase in the supply of funds), which mechanically leads to a decrease in utilization. Conversely, a decrease in the interest rate leads to an increase in borrowing demand (and correspondingly a reduction in the supply of funds), resulting in an increase in utilization. This model is designed to capture the microstructure of the liquidity pool and is therefore valid only locally in time. Calibration to block-by-block market data is presented in Section \ref{section:data_processing}.

The wealth generated for lenders depends on the interest paid by borrowers. However, the total interest $r_{t} B_{t}$ paid by borrowers must be divided by $S_{t}$, as these payments are distributed among lenders in proportion to their individual contributions to the pool. Consequently, given \eqref{eq:utilization_rate}, the lender wealth, denoted by $X_t$, evolves according to the following dynamics:
\begin{equation}\label{eq:wealth_dynamics}
dX_{t} = r_{t} U_{t} dt.
\end{equation}
Inherently, wealth should be compounded as outlined in \citet{whitepaper2020aavev1}. However, given the very short time horizon considered in this study, compounding is neglected.

We aim to determine, in real-time, the interest rate model that maximizes lender wealth generated over a time horizon $T$, while:
\begin{itemize}
    \item stabilizing the interest rate over time through a running penalty term $\phi (r_{t} - \bar{r})^{2}$, where $\bar{r} \geq 0 $ is a predetermined reference rate and $\phi > 0$ is the rate stabilization parameter of the protocol;
    \item mitigating the terminal liquidity risk (i.e., the utilization rate at time $T$ is close to $1$) using a penalty function $\psi$. For a given target utilization level, denoted by $u^{*} \in [0,1]$, the penalty function, defined for $u \in [0,1]$, is given by:
    \begin{equation}\label{eq:terminal_penalty_function}
        \psi(u) = \eta \big [ \max (u-u^{*}, 0) \big ]^{2},
    \end{equation}
    where the parameter $\eta \geq 0$ is the liquidity risk aversion parameter of the protocol. Lending protocols commonly use a target utilization rate parameter to define the interest rate model \citep{whitepaper2020aavev1}. Accordingly, our approach introduces a quadratic penalty for utilization levels that exceed this target rate. Moreover, this penalty is applied at the terminal horizon rather than continuously, as the time frame considered in this paper is very short.
\end{itemize}

The combination of expected lender wealth maximization, rate stabilization and liquidity risk mitigation leads to the following stochastic control problem:
\begin{equation}\label{eq:control_problem}
\sup_{(r_{t})_{t \geq 0} \in \mathcal{A}} \mathbb{E} \Big [ X_{T} - \psi(U_{T}) - \phi \int_{0}^{T} (r_{t} - \bar{r})^{2} dt \Big ],
\end{equation}
where $X_{T}$ is the wealth of the liquidity pool at time $T$ and $U_{T}$ is the utilization rate at time $T$.

The objective function in \eqref{eq:control_problem} is designed to benefit both lenders and borrowers. On the one hand, it aims to maximize the expected lender wealth and incorporates a penalty for liquidity risk controlled by the parameter $\eta$. On the other hand, borrower activity is indirectly encouraged since lender yield depends on the utilization rate. In addition, the interest rate stabilization penalty, governed by the parameter $\phi$, prevents excessive borrowing costs. If used by a protocol, both parameters must be set to reflect a collective level of risk aversion shared by lenders and borrowers. In practice, risk parameters are set by protocol governance; see, for instance, \citet{whitepaper2020aavev1}. Changes are submitted to vote and voting power is proportional to the quantity of governance tokens held. Importantly, token holders may be either lenders or borrowers, who have opposing interests: lenders seek high interest rates, whereas borrowers prefer low ones. Since both risk parameters influence interest rates, assigning control of $\eta$ to lenders and $\phi$ to borrowers is not a viable option. Hence, governance outcomes must reflect a majority consensus among token holders, despite opposing interests and differing views on risk.

The problem \eqref{eq:control_problem} does not admit a closed-form optimal solution in the case of general intensities. In Section \ref{section:linear_intensities}, the intensities are assumed to be linear functions of the interest rate. The optimal interest rate model is then derived from a system of Riccati-type ODEs, which can be solved numerically. In Section \ref{section:non_linear_intensities}, the linearity assumption is relaxed and a Monte-Carlo estimator coupled with deep learning techniques is employed to approximate the optimal interest rate model. Finally, in Section \ref{section:parametric_rates}, we consider industry-standard parametric interest rate models and propose a method for estimating the parameters governing these models.

\subsection{Linear intensities}\label{section:linear_intensities}

We assume that the intensities are linear functions of the interest rate:
\begin{equation}\label{eq:linear_intensities}
\lambda^{+}(r) =  a_{0}^{+} + a_{1}^{+} r, \quad \lambda^{-}(r) = a_{0}^{-} + a_{1}^{-} r,
\end{equation}
where $a_{0}^{+} \geq 0$, $a_{0}^{-} \geq 0$, $a_{1}^{+} \leq 0$ and $a_{1}^{-} \geq 0$.

Following \citet[Section 5.4.2; Equation (5.39)]{cartea2015algorithmic}, the problem \eqref{eq:control_problem} is characterized by a value function $(x, u, t) \mapsto v(x, u, t)$ and the associated Hamilton-Jacobi-Bellman (HJB) equations with $u \in \{ \delta, \ldots, 1 - \delta\}$ and $(x, t) \in \mathbb{R} \times \lbrack 0, T \rbrack $:
\begin{equation}\label{eq:ode_first}
\begin{split}
\frac{\partial v}{\partial t} + \sup_{r} \Big ( ru \frac{\partial v}{\partial x} & + \lambda^{+}(r) [v(x, u + \delta, t) - v(x, u, t)] \\
& + \lambda^{-}(r) [v(x, u - \delta, t) - v(x, u, t)] - \phi (r - \bar{r})^{2} \Big ) = 0,
\end{split}
\end{equation}
with the terminal condition:
\begin{equation}\label{eq:terminal_ode_first}
v(x, u, T) = x - \psi(u).
\end{equation}

In light of the non-compounding nature of the wealth dynamics, we consider the following \textit{ansatz}:
\begin{equation}\label{eq:ansatz}
v(x, u, t) = x + h(u, t),
\end{equation}
for some function $h$ with $u \in \{ \delta, \ldots, 1 - \delta\}$ and $t \in \lbrack 0, T \rbrack$.
By plugging the \textit{ansatz} \eqref{eq:ansatz} into \eqref{eq:ode_first} and \eqref{eq:terminal_ode_first}, the problem reduces to two dimensions:
\begin{equation}\label{eq:ode_ansatz}
\begin{split}
\frac{\partial h}{\partial t} + \sup_{r} \Big (ru & + \lambda^{+}(r) [h(u + \delta, t) - h(u, t)] \\
& + \lambda^{-}(r) [h(u - \delta, t) - h(u, t)] - \phi (r - \bar{r})^{2} \Big ) = 0,
\end{split}
\end{equation}
with the terminal condition:
\begin{equation}\label{eq:terminal_ode_ansatz}
h(u, T) = - \psi(u).
\end{equation}

By plugging equations \eqref{eq:linear_intensities} into \eqref{eq:ode_ansatz} and \eqref{eq:terminal_ode_ansatz} and taking the supremum, the optimal interest rate model reads:
\begin{equation}\label{eq:optimal_rate}
r^{*}(u, t) = \bar{r} + \frac{1}{2 \phi} \Big [ u + a_{1}^{+} h(u + \delta, t) + a_{1}^{-} h(u - \delta, t) - (a_{1}^{+} + a_{1}^{-}) h(u, t) \Big ],
\end{equation}
with
\begin{equation}\label{eq:ode_riccati}
\begin{split}
\frac{\partial h}{\partial t} + \bar{r} u
& + \lambda^{+} (\bar{r}) [h(u+\delta, t) - h(u, t)] + \lambda^{-} (\bar{r}) [h(u-\delta, t) - h(u, t)] \\
& + \frac{1}{4\phi} \Big [ u + a_{1}^{+} h(u+\delta, t) + a_{1}^{-} h(u-\delta, t) - (a_{1}^{+} + a_{1}^{-}) h(u, t) \Big ]^{2} = 0,
\end{split}
\end{equation}
and the terminal condition:
\begin{equation}\label{eq:terminal_ode_riccati}
h(u, T) = -\psi(u).
\end{equation}
Equations \eqref{eq:ode_riccati} and \eqref{eq:terminal_ode_riccati} define a system of ODEs of Riccati type for $u \in \{ \delta, \ldots, 1 - \delta\}$ and $t \in \lbrack 0, T \rbrack$. Moreover, we need to introduce boundary conditions to determine $h (0, t)$ and $h (1, t)$ for $t \in \lbrack 0, T \lbrack$. Motivated by the terminal penalty function \eqref{eq:terminal_penalty_function}, the boundary conditions considered impose that the third derivatives, in their discretized form, equal zero:
\begin{equation}\label{eq:boundary_condition_1}
h (0, t) = 3 h (\delta, t) - 3 h (2 \delta, t) + h (3 \delta, t),
\end{equation}
\begin{equation}\label{eq:boundary_condition_2}
h (1, t) = 3 h (1 - \delta, t) - 3 h (1 - 2 \delta, t) + h (1 - 3 \delta, t).
\end{equation}

Let us define $h(t) = \big ( h(0, t), h(\delta, t), \ldots, h(1-\delta, t), h(1, t)
\big )^{\top}$ and $U = \big ( 0, \delta, \ldots, 1 - \delta, 1 \big )^{\top}$. Using $\odot$ the element-wise (Hadamard) product, the problem can be reformulated as follows:
\begin{equation}\label{eq:ode_riccati_vector}
\begin{split}
\frac{\partial h}{\partial t} + \bar{r} U + A h(t) + \frac{1}{4\phi} \Big ( U + B h(t) \Big ) \odot \Big ( U + B h(t) \Big ) = 0,
\end{split}
\end{equation}
with the terminal condition:
\begin{equation}\label{eq:terminal_ode_riccati_vector}
h(T) = -\psi(U),
\end{equation}
and
\begin{equation}\label{eq:matrix_a_b}
\begin{split}
A & = \begin{pmatrix}
2 \bar{\lambda}^{-} - \bar{\lambda}^{+} & \bar{\lambda}^{+} - 3 \bar{\lambda}^{-} & \bar{\lambda}^{-} \\
\bar{\lambda}^{-} & - \bar{\lambda}^{-} - \bar{\lambda}^{+} & \bar{\lambda}^{+} & & (0) \\
& \ddots & \ddots & \ddots \\
(0) & & \bar{\lambda}^{-} & - \bar{\lambda}^{-} - \bar{\lambda}^{+} & \bar{\lambda}^{+} \\
& & \bar{\lambda}^{+} & \bar{\lambda}^{-} - 3 \bar{\lambda}^{+} & - \bar{\lambda}^{-} + 2 \bar{\lambda}^{+}
\end{pmatrix},\\
\\
B & = \begin{pmatrix}
2 a_{1}^{-} - a_{1}^{+} & a_{1}^{+} - 3 a_{1}^{-} & a_{1}^{-}\\
a_{1}^{-} & - a_{1}^{-} - a_{1}^{+} & a_{1}^{+} & & (0) \\
& \ddots & \ddots & \ddots \\
(0) & & a_{1}^{-} & - a_{1}^{-} - a_{1}^{+} & a_{1}^{+} \\
& & a_{1}^{+} & a_{1}^{-} - 3 a_{1}^{+} & - a_{1}^{-} + 2 a_{1}^{+}
\end{pmatrix}.
\end{split}
\end{equation}
The solution to the system \eqref{eq:ode_riccati_vector} with \eqref{eq:terminal_ode_riccati_vector} can be approximated numerically.

\subsection{Nonlinear intensities: a neural network technique}\label{section:non_linear_intensities}

In this section, we present a Monte-Carlo approach, coupled with deep learning techniques, to approximate the optimal interest rate model for arbitrary intensity functions. We parameterize the interest rate model using a feedforward neural network and the loss function used for training is derived from the objective function in \eqref{eq:control_problem}. We start by adopting a regular partition of the time interval $[0, T]$: $0 = t_{0} < t_{1} < \ldots < t_{N} = T$. The time step is denoted by $\tau = \frac{T}{N}$. The interest rate model is parameterized via a feedforward neural network, denoted by $\hat{r}$, with a set of parameters $\hat{\theta}$.

From the dynamics of the utilization rate \eqref{eq:utilization_dynamics}, it follows that for $i \in \{ 0, \ldots, N-1 \}$:
\begin{equation}\label{eq:Euler_utilization}
U_{t_{i+1}} = U_{t_{i}} + \delta \Delta N^{+}_{t_{i}} - \delta \Delta N^{-}_{t_{i}},
\end{equation}
with $\Delta N^{\pm}_{t_{i}} = N^{\pm}_{t_{i+1}} - N^{\pm}_{t_{i}}$ and $U_{0} = u_{0}$.

$\Delta N^{\pm}_{t_{i}}$ can be approximated by Binomial random variables with parameters $J \in \mathbb{N}^{*}$ and probabilities $p^{\pm}_{t_{i}} = \lambda^{\pm} \big ( \hat{r}(u, t_{i}) \big ) \tau J^{-1}$ for a given utilization rate $u$, assuming $J$ is sufficiently large. In this case, we have:
\begin{equation}\label{eq:jump_naive}
\Delta N^{\pm}_{t_{i}} \approx \sum_{j=0}^{J-1} H (p^{\pm}_{t_{i}} - Z^{j, \pm}_{t_{i}}),
\end{equation}
where $Z^{j, \pm}_{t_{i}}$ are uniformly distributed over the interval $(0, 1)$ and $H(x) = \mathbbm{1}_{x \geq 0}$ is the Heaviside function. The Heaviside function has derivatives equal to zero everywhere, except at $x=0$ where it is infinite. Consequently, gradient-based optimization cannot be applied, prohibiting the feedforward neural network from learning the optimal solution. To circumvent this issue, we use a continuous relaxation of the Bernoulli distribution, known as the reparametrization trick and introduced in \citet{maddison2016concrete}:
\begin{equation}\label{eq:jump_smoothed}
\Delta N^{\pm}_{t_{i}} \approx \sum_{j=0}^{J-1} H^{\varepsilon} \big ( L ( p^{\pm}_{t_{i}} ) + L ( Z^{j, \pm}_{t_{i}} ) \big ),
\end{equation}
where $H^{\varepsilon}(x) = \frac{\max (x+\varepsilon, 0) - \max (x-\varepsilon, 0)}{2 \varepsilon}$ and $L(x) = \log (x) - \log (1-x)$. In \citet{maddison2016concrete}, the sigmoid function is used to smooth the Heaviside function. However, we opt for the hard-sigmoid function as it produces more stable numerical results.

From \eqref{eq:wealth_dynamics}, the Euler scheme for the wealth process is:
\begin{equation}\label{eq:Euler_wealth}
X_{t_{i+1}} = X_{t_{i}} + \hat{r}(U_{t_{i}}, t_{i}) U_{t_{i}} \tau,
\end{equation}
with $X_{0} = 0$.

Hence, we maximize over $\hat{\theta}$ the discretized form of the objective function in \eqref{eq:control_problem}, denoted by $L$:
\begin{equation}\label{eq:loss_function}
L(\hat{\theta}) = \mathbb{E} \Big [ X_{t_{N}} - \psi(U_{t_{N}}) - \phi \sum_{i=0}^{N-1} \big ( \hat{r}(U_{t_{i}}, t_{i}) - \bar{r} \big )^{2} \tau \Big ].
\end{equation}
Additionally, we incorporate a penalty function, denoted by $P$, to enforce convexity with respect to the utilization rate at each time step, which reads:
\begin{equation}\label{eq:penalty_function}
P(\hat{\theta}) = \sum_{i=0}^{N-1} \sum_{u \in \{ u^{*}, \ldots, 1 - \delta \}} \min \big ( \hat{r}(u + \delta, t_{i}) - 2 \hat{r}(u, t_{i}) + \hat{r}(u - \delta, t_{i}), 0 \big ).
\end{equation}
The total loss function used to train the feedforward neural network is composed of the objective function $L$ along with the additional penalty term $P$. In \citet{chataigner2020deep}, the authors propose two approaches to ensure that the neural network is a convex function of a given input. The first approach involves adapting the architecture of the neural network to embed shape conditions as hard constraints. The second one employs shape penalization to favor these conditions as soft constraints. Following the recommendations of \citet{chataigner2020deep}, we adopt the soft constraint technique through the penalty function $P$.

Note that, due to the use of a neural network, there is no guarantee of convergence to the global optimum of the stochastic control problem \eqref{eq:control_problem}. To validate our training methodology empirically, Appendix \ref{appendix:consistency_checks} presents consistency checks in the case of linear intensities.

This methodology allows us to approximate the optimal interest rate model for arbitrary intensity functions, thereby relaxing the linearity assumption considered in Section \ref{section:linear_intensities}. However, the models obtained here and in Section \ref{section:linear_intensities} share structural limitations. Although they depend on the current utilization ratio, as is common in the industry, they are also time-dependent which limits their practicality. Implementing such models would require setting a fixed maturity and continually rolling it forward. In addition, deriving the models necessitates either numerically solving a system of ODEs or training a neural network, both of which are computationally intensive, particularly the second. For these reasons, Section \ref{section:parametric_rates} turns to industry-standard parametric models and estimates their optimal parameters. Here, optimality means that the discretized form of the objective function in \eqref{eq:control_problem} is maximized.

\subsection{Parametric models}\label{section:parametric_rates}

We replace the neural network used in Section \ref{section:non_linear_intensities} with a parametric interest rate model, denoted by $\tilde{r}$, characterized by a set of parameters $\tilde{\theta}$. The Monte-Carlo framework, as described in Section \ref{section:non_linear_intensities}, remains unchanged and we determine the optimal parameters such that the function \eqref{eq:loss_function} is maximized. In contrast to Section \ref{section:non_linear_intensities}, no penalty function is used in the loss function (i.e., $P(\tilde{\theta}) = 0$) as the convex constraint is enforced through penalty on the parameters. Details on the parameter penalty are presented in Section \ref{section:architecture}.

\citet{gudgeon2020defi} conducts a comprehensive review of the most prevalent interest rate models used in the industry, with a focus on the linear and kinked rates. The kinked rate refers to the interest rate model introduced in \citet{whitepaper2020aavev1}. Both models belong to the first generation of interest rate models, which only depend on the current utilization rate of the liquidity pool. More recently, a new generation of models has emerged, notably with the introduction of the Morpho interest rate model \citep{adaptive2023morpho}, called AdaptiveCurveIRM (abbreviated as adaptive rate). The structure of this interest rate model is dynamic and designed to enhance market efficiency by targeting higher capital utilization and improving risk management, as evidenced in \citet{bertucci2024agents}. We study the three models:
\begin{itemize}
    \item Linear rate: The linear interest rate model is the simplest formulation and is given by:
    \begin{equation}
    r(u) = r_{base} + \frac{u}{u^{*}} r_{slope1},
    \end{equation}
    where $\tilde{\theta} = \big ( r_{base}, r_{slope1} \big )$, $r_{base} \geq 0$ and $r_{slope1} \geq 0$.
    \item Kinked rate: The kinked interest rate model is expressed as:
    \begin{equation}
    r(u) = \left\{
        \begin{array}{ll}
            r_{base} + \frac{u}{u^{*}} r_{slope1} & \mbox{if } u < u^{*} \\
            r_{base} + r_{slope1} + \frac{u - u^{*}}{1 - u^{*}} r_{slope2} & \mbox{if } u \geq u^{*},
        \end{array}
    \right.
    \end{equation}
    where $\tilde{\theta} = \big ( r_{base}, r_{slope1}, r_{slope2} \big )$, $r_{base} \geq 0$, $r_{slope1} \geq 0$ and $r_{slope2} \geq 0$.
    \item Adaptive rate: The adaptive rate depends on the current utilization rate and the previous state of the liquidity pool. We denote by $t_{last}$ the last time the pool is modified and by $u_{t_{last}}$ the utilization rate of the pool at that time. The model is then given by:    
    \begin{equation}
    r(u,t) = r^{\text{target}}_{t} \text{curve}(u),
    \end{equation}
    where,
    \begin{equation}
    r^{\text{target}}_{t} = r^{\text{target}}_{t_{last}} \text{speed}(t),
    \end{equation}
    \begin{equation}
    \text{speed}(t) = e^{k_{p} \text{error}(u_{t_{last}}) (t - t_{last})}, \quad 
    \text{error}(u) = \left\{
    \begin{array}{ll}
        \frac{u - u^{*}}{u^{*}} & \mbox{if } u < u^{*} \\
        \frac{u - u^{*}}{1 - u^{*}} & \mbox{if } u \geq u^{*},
    \end{array}
    \right.
    \end{equation}
    and,
    \begin{equation}
    \text{curve}(u) = \left\{
        \begin{array}{ll}
            \big ( 1 - k_{d_{1}} \big ) \text{error}(u) + 1 & \mbox{if } u < u^{*} \\
            \big ( k_{d_{2}} - 1 \big ) \text{error}(u) + 1 & \mbox{if } u \geq u^{*},
        \end{array}
    \right.
    \end{equation}
    where $\tilde{\theta} = \big ( r^{\text{target}}_{0}, k_{p}, k_{d_{1}}, k_{d_{2}} \big )$, $r^{\text{target}}_{0} \geq 0$, $k_{p} \geq 0$, $k_{d_{1}} \geq 0$ and $k_{d_{2}} \geq 0$.
\end{itemize}

Regarding the linear and kinked models, we use the notations introduced in \citet{whitepaper2020aavev1}. Specifically, the parameter $r_{slope1}$ has been normalized such that both rates at target utilization equals $r_{base} + r_{slope1}$. Additionally, for the kinked rate model, the parameter $r_{slope2}$ has also been normalized such that the interest rate at full utilization equals $r_{base} + r_{slope1} + r_{slope2}$.

Although the three models depend on the current utilization of the liquidity pool, the adaptive model also incorporates time dependence via the parameter $r^{\text{target}}_{t}$. Moreover, each model uses a target utilization rate, denoted by $u^{*}$, presumed to be already set by the platform and therefore excluded from $\tilde{\theta}$.

\section{Numerical results using synthetic linear intensities}\label{section:numerical_results}

\subsection{Illustration}\label{section:toy_example}

We begin by considering a scenario with synthetic data to illustrate the behavior of the optimal interest rate with respect to both time and the utilization rate. The parameters used are: $T = 100$ blocks, $\delta = 0.01$, $u^{*} = 0.9$, $\phi = 7$, $\eta = 1500$ and $\bar{r} = 0$. The time horizon $T$ is expressed in terms of blocks. On the Ethereum blockchain, a block is typically generated every $12$ seconds, though this duration may vary because of network congestion. Additionally, synthetic linear intensity functions are employed.
\begin{figure}
    \centering
    \includegraphics[scale = 0.6]{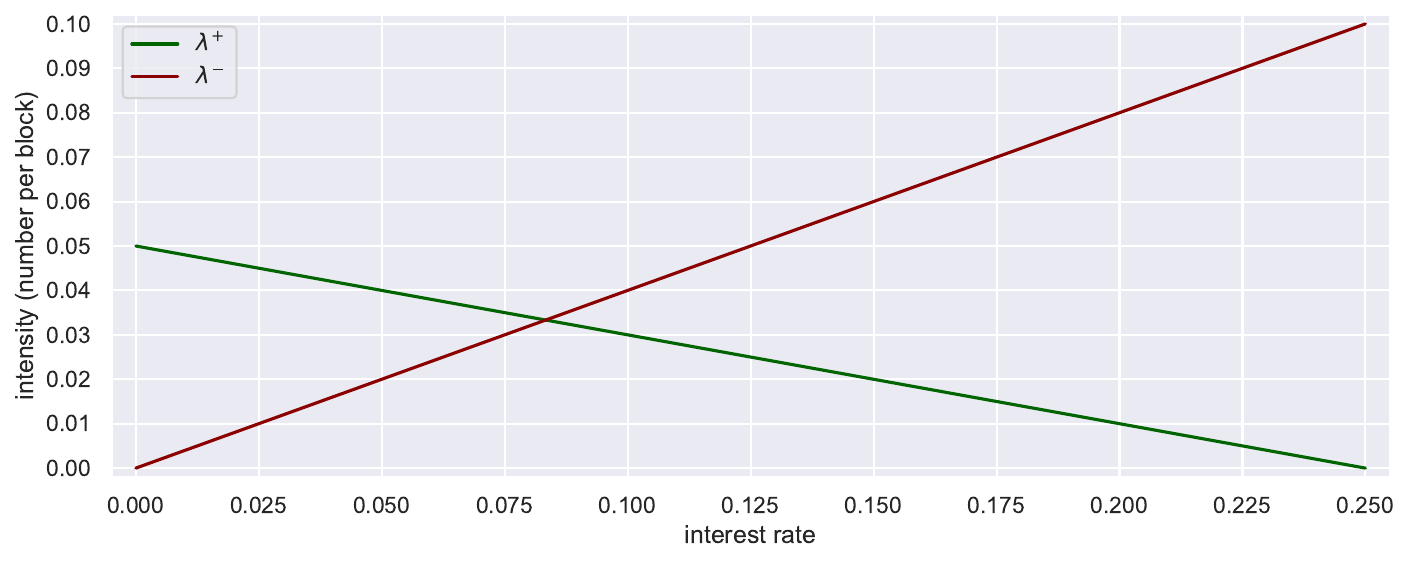}
    \caption{Synthetic linear intensity functions with respect to the interest rate, as defined in \eqref{eq:linear_intensities}; $a_{0}^{+} = 0.05$, $a_{1}^{+} = - 0.2$, $a_{0}^{-} = 0$ and $a_{1}^{-} = 0.4$.}    
    \label{fig:toy_example_intensities}
\end{figure}
Figure \ref{fig:toy_example_intensities} shows the intensity functions, representing the arrival rates of events that respectively increase or decrease the utilization rate by $\delta$ given the interest rate. The function $\lambda^{+}$ is intentionally configured to decrease linearly as the interest rate rises. This design choice reflects the disincentive effect of higher borrowing costs, as defined by our parameters. Specifically, as the interest rate rises, the cost of taking out new loans becomes more prohibitive for potential borrowers, leading to a reduction in the number of events that would increase the utilization rate. Conversely, the function $\lambda^{-}$ is specified to increase linearly as the interest rate rises. This modeling choice aligns with the rational hypothesis that as borrowing costs increase, borrowers are more inclined to repay their loans to avoid the higher interest expenses, leading to more events that decrease the utilization rate. From the perspective of liquidity providers, higher interest rates may serve as a strong incentive to boost deposits into the liquidity pool. Finally, the intersection of $\lambda^{+}$ and $\lambda^{-}$ at $0.0833$ reflects an equilibrium interest rate in the market dynamics, where the pressure to either increase or decrease utilization is balanced.

\begin{figure}
    \centering
    \includegraphics[scale = 0.6]{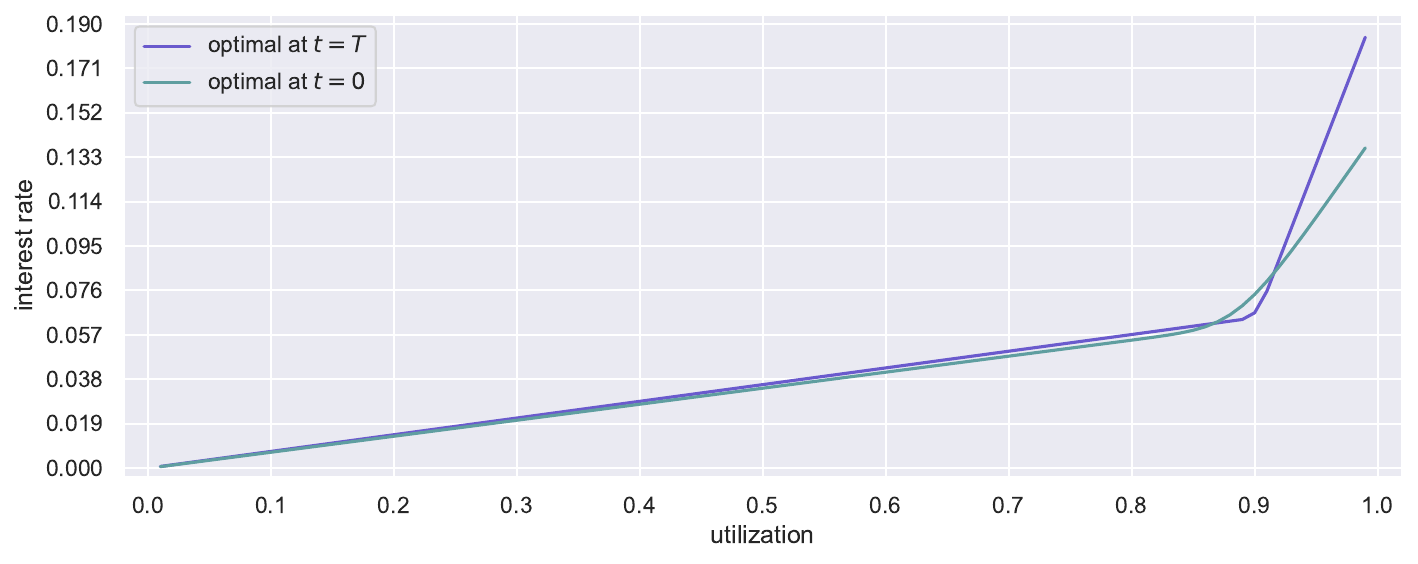}
    \caption{Optimal interest rate as a function of utilization at $t = T$ and $t = 0$, obtained through the numerical approximation of the system of ODEs; $T = 100$ blocks, $\delta = 0.01$, $u^{*} = 0.9$, $\phi = 7$, $\eta = 1500$ and $\bar{r} = 0$.}
    \label{fig:toy_example_rates}
\end{figure}

The synthetic intensity functions have been chosen to be linear in order to adhere to the assumptions made in Section \ref{section:linear_intensities}. Thus, we approximate the optimal interest rate model \eqref{eq:optimal_rate} by solving the system of ODEs \eqref{eq:ode_riccati_vector} and \eqref{eq:terminal_ode_riccati_vector} numerically using the SciPy \textit{odeint} function.\footnote{\url{https://docs.scipy.org/doc/scipy/reference/generated/scipy.integrate.odeint.html}} With a slight abuse of terminology, we refer to the numerical approximation as the optimal interest rate model. Figure \ref{fig:toy_example_rates} illustrates it with respect to the utilization rate at the initial time ($t = 0$) and at maturity ($t = T$).
The optimal interest rate curve at $T$ is a piecewise-linear function of the utilization, with a significant increase in slope beyond the target utilization rate $u^{*}$. Indeed, by performing a Taylor expansion of the optimal interest rate \eqref{eq:optimal_rate} at maturity for $\delta$ close to $0$, we obtain:
\begin{equation}\label{eq:Taylor_expansion}
r^{*}(u, T) \approx \bar{r} + \frac{1}{2 \phi} \Big [ u + 2 \eta \delta (a_{1}^{-} - a_{1}^{+}) \max (u-u^{*}, 0) \Big ].
\end{equation}
Therefore, for very short maturities and linear intensity functions, the optimal interest rate curve is piecewise-linear with two distinct slopes and corresponds to the kinked model. The initial point of the curve at $0$ is given by $\bar{r}$. The slope of the first segment of the curve (between $0$ and $u^{*}$) is exclusively determined by the parameter $\phi$, while beyond $u^{*}$, the slope is governed by the parameters $a_{1}^{+}$, $a_{1}^{-}$, $\delta$ and $\eta$, along with $\phi$. Additionally, both slopes are positive. From a financial perspective, the first segment of the curve at maturity reflects a relatively stable borrowing environment where the interest rate responds proportionally and moderately to increased utilization. In contrast, the second segment of the interest rate curve demonstrates a sharp escalation, marking a significant shift in the interest rate behavior. As the pool approaches full utilization, this steep incline represents a significant penalty for borrowing, driven by the scarcity of available liquidity. In this situation, the model is designed to rapidly increase borrowing costs to discourage further activities and prevent the pool from reaching full capacity.

The optimal initial interest rate curve is flatter than the terminal curve because of the stabilization penalty term in \eqref{eq:control_problem}. For $u \leq u^{*}$, the initial and terminal interest rates are close. In contrast, for $u > u^{*}$, the initial interest rate is lower than the terminal one. This behavior can be interpreted as a measure to avoid discouraging early borrowing, maintaining sufficient utilization to maximize wealth in the initial stages. As time progresses, the interest rate increases steeply to manage the terminal liquidity risk and align with the long-term objectives of the protocol. The model ensures the protocol balances short-term activity and long-term liquidity management effectively.

In Appendix \ref{appendix:consistency_checks}, we compare the ODEs-based interest rate model \eqref{eq:optimal_rate} to the deep learning-based one from Section \ref{section:non_linear_intensities}. The results demonstrate that the neural network-based methodology provides accurate estimates of the optimal interest rate model in the case of linear intensities.

\subsection{Sensitivity analysis}\label{section:influence_parameters}

\begin{figure}
    \centering
    \begin{subfigure}{\linewidth} 
        \centering
        \includegraphics[scale = 0.6]{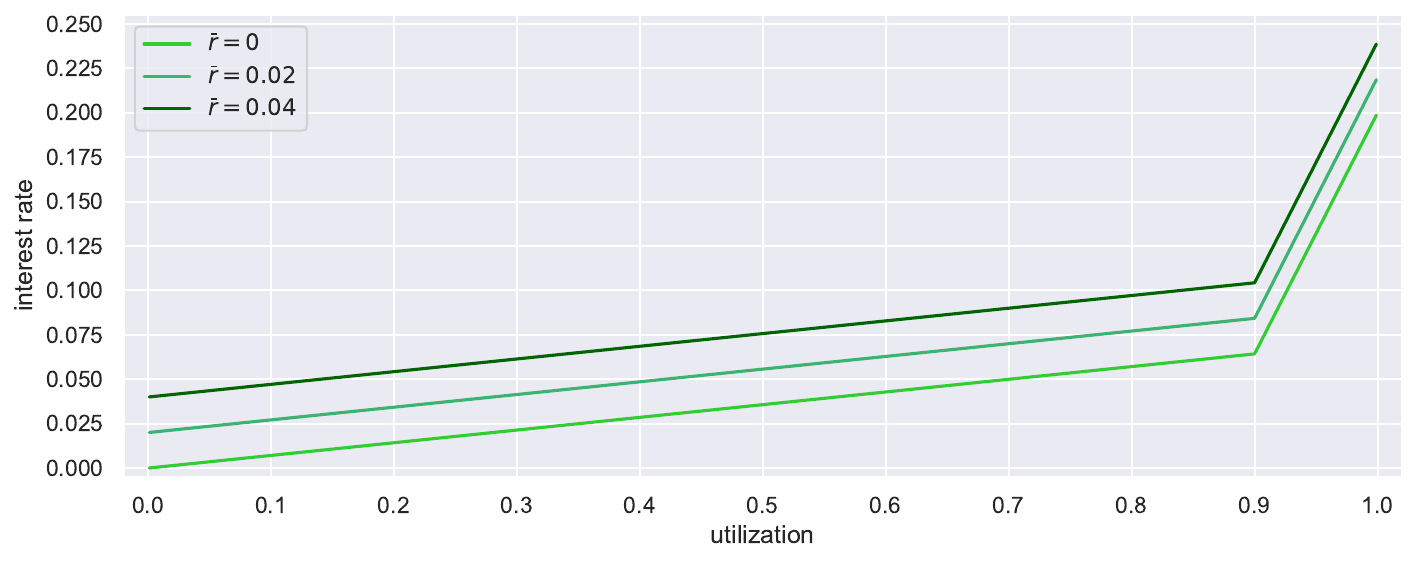}
        \caption{Influence of $\bar{r}$; $\phi = 7$ and $\eta = 1500$}
        \label{fig:r_bar_influence}
    \end{subfigure}
    \vfill
    \begin{subfigure}{\linewidth}
        \centering
        \includegraphics[scale = 0.6]{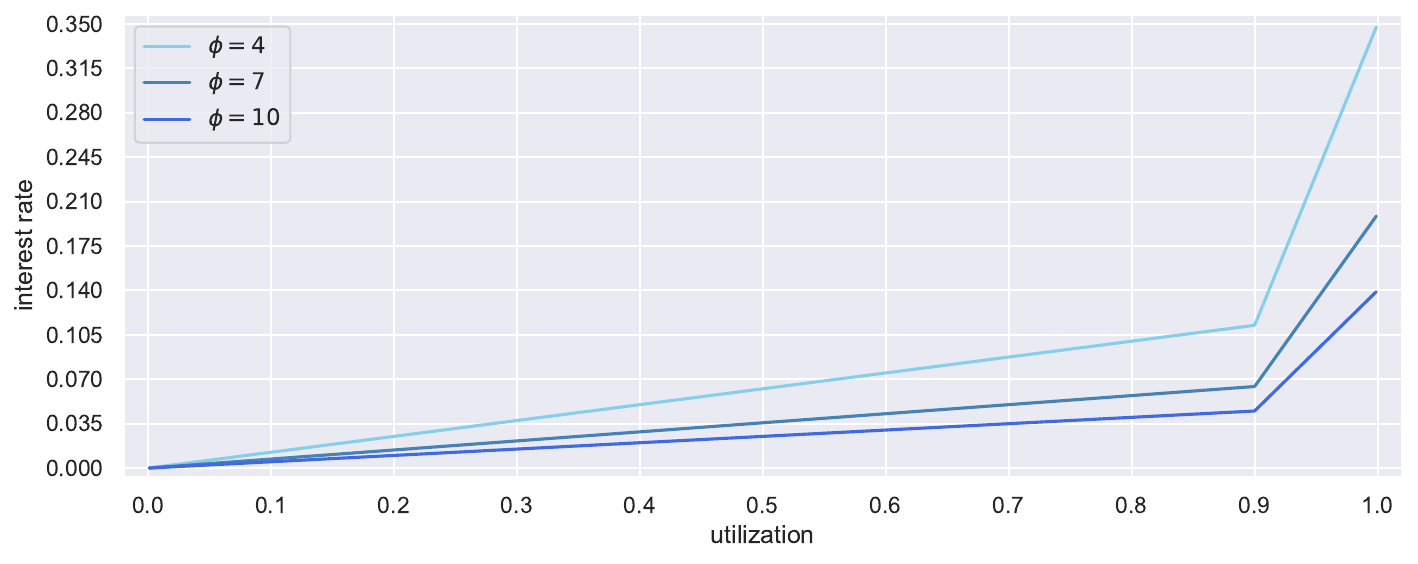}
        \caption{Influence of $\phi$; $\eta = 1500$ and $\bar{r} = 0$}
        \label{fig:phi_influence}
    \end{subfigure}
    \vfill
    \begin{subfigure}{\linewidth}
        \centering
        \includegraphics[scale = 0.6]{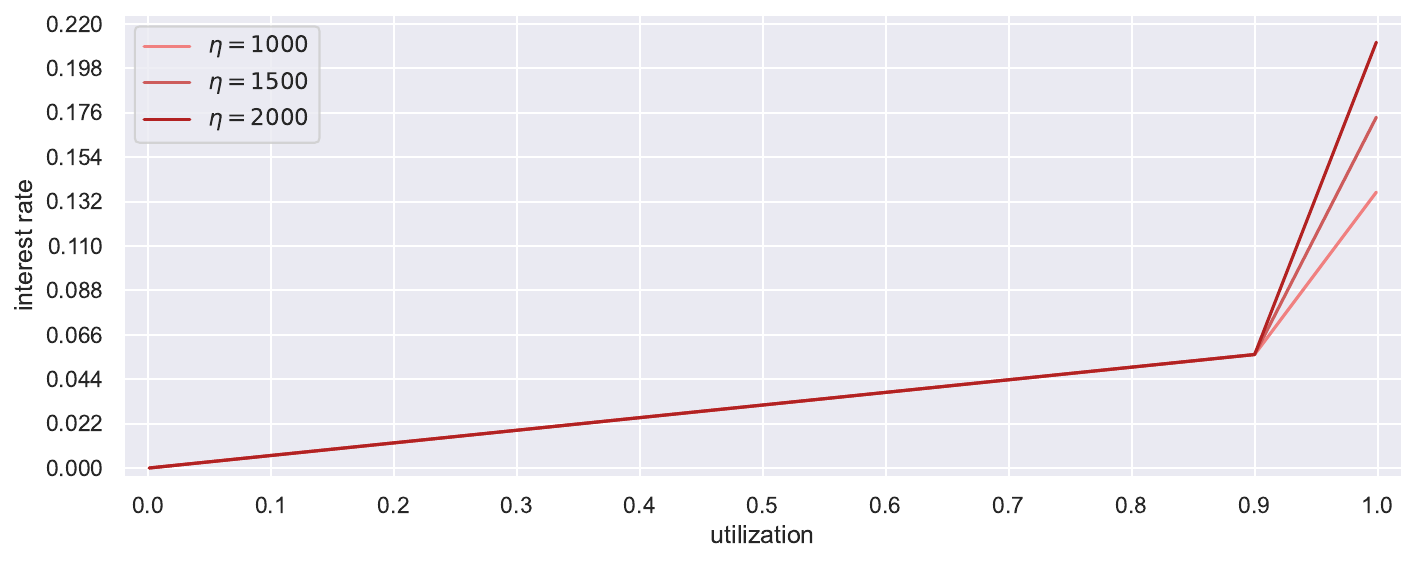}
        \caption{Influence of $\eta$; $\phi = 7$ and $\bar{r} = 0$}
        \label{fig:eta_influence}
    \end{subfigure}
    \caption{Influence of parameters $\bar{r}$, $\phi$ and $\eta$ on the optimal terminal interest rate as a function of utilization; $T = 100$ blocks, $\delta = 0.01$, $u^{*} = 0.9$.}    
    \label{fig:parameters_influence}
\end{figure}
This section investigates the influence of the risk parameters on the optimal terminal interest rate curve. Although this analysis is limited to the terminal rate, it provides valuable insights into how different parameters shape the interest rate model introduced in Section \ref{section:linear_intensities}. We begin with the reference rate $\bar{r}$, which governs the overall level of the curve and sets the interest rate when utilization is at $0$ (see Figure \ref{fig:r_bar_influence}). Additionally, it follows from \eqref{eq:optimal_rate} that, for $u \in [0,1]$:
\begin{equation}\label{eq:r_bar_influence}
\frac{\partial r^{*}(u, T)}{\partial \bar{r}} = 1.
\end{equation}

Parameter $\phi$ controls the running penalty on the interest rate stabilization. Hence, from equation \eqref{eq:Taylor_expansion}, we have for $u \in [0,1]$:
\begin{equation}\label{eq:phi_influence}
\frac{\partial r^{*}(u, T)}{\partial \phi} \approx - \frac{1}{2 \phi^{2}} \Big [ u + 2 \eta \delta (a_{1}^{-} - a_{1}^{+}) \max (u-u^{*}, 0) \Big ] \leq 0.
\end{equation}
Figure \ref{fig:phi_influence} illustrates the influence of $\phi$. As it increases, the model tends to flatten the optimal interest rate curve towards $\bar{r}$.

Parameter $\eta$ controls the terminal penalty on the utilization rate. Thus, its influence is limited to the second segment of the optimal interest rate curve beyond $u^{*}$ because of the nature of the terminal penalty \eqref{eq:terminal_penalty_function}. Hence, we have for $u \in [0,1]$:
\begin{equation}\label{eq:eta_influence}
\frac{\partial r^{*}(u, T)}{\partial \eta} \approx \frac{1}{\phi} \Big [ \delta (a_{1}^{-} - a_{1}^{+}) \max (u-u^{*}, 0) \Big ] \geq 0.
\end{equation}
As the influence of the terminal penalty on the utilization rate increases, the model steepens the interest rate curve beyond $u^{*}$ to foster activity that reduces utilization, thereby mitigating liquidity risk. This behavior is illustrated in Figure \ref{fig:eta_influence}.

\section{Application to lending protocols}\label{section:applications}

\subsection{Data processing and intensity calibration}\label{section:data_processing}

\begin{figure}
    \centering
    \begin{subfigure}{\linewidth} 
        \centering
        \includegraphics[scale = 0.6]{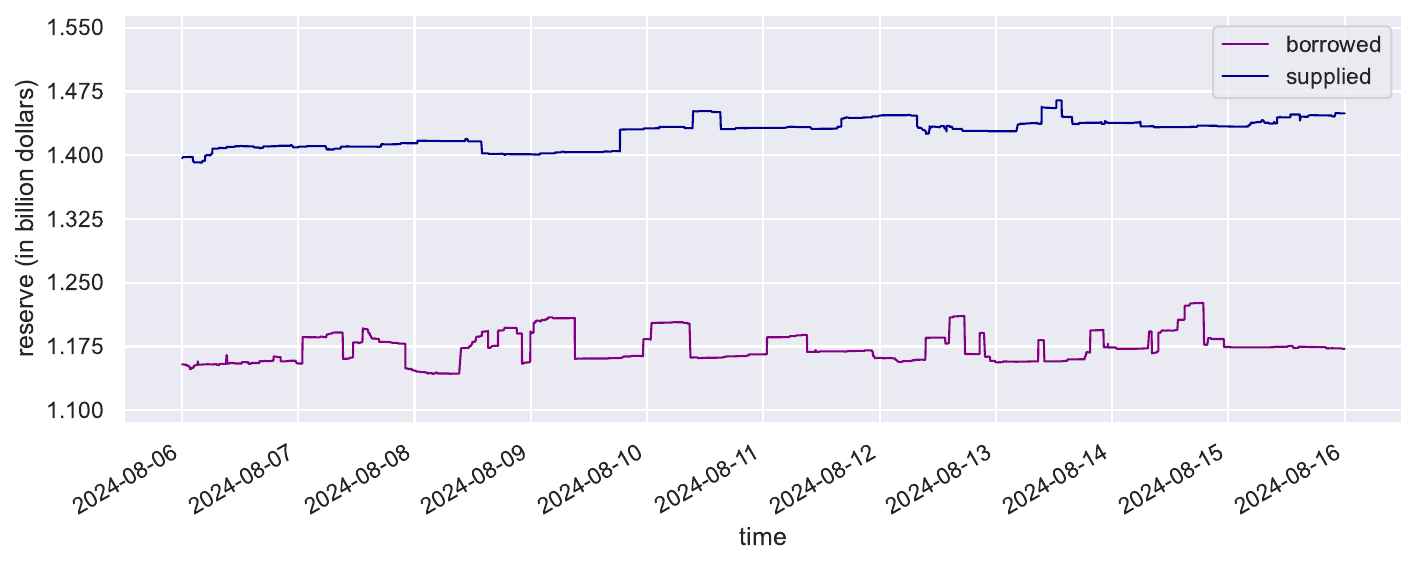}
        \caption{Reserves}
        \label{fig:reserves}
    \end{subfigure}
    \vfill
    \begin{subfigure}{\linewidth}
        \centering
        \includegraphics[scale = 0.6]{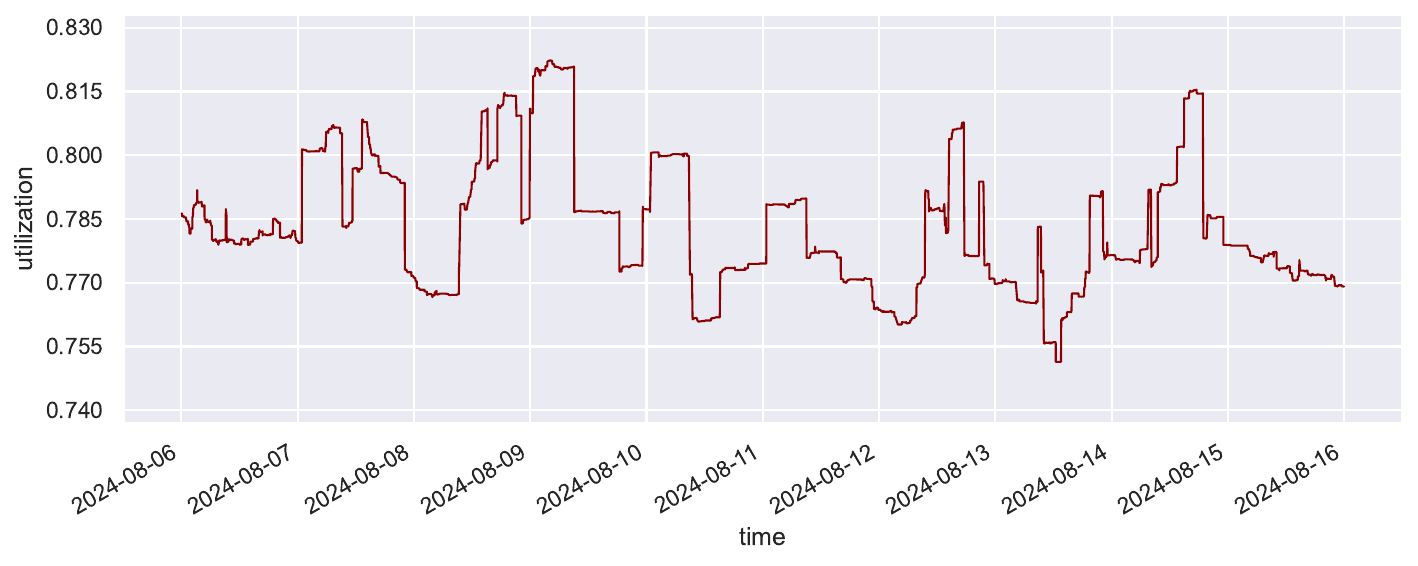}
        \caption{Utilization}
        \label{fig:utilization}
    \end{subfigure}
    \vfill
    \begin{subfigure}{\linewidth}
        \centering
        \includegraphics[scale = 0.6]{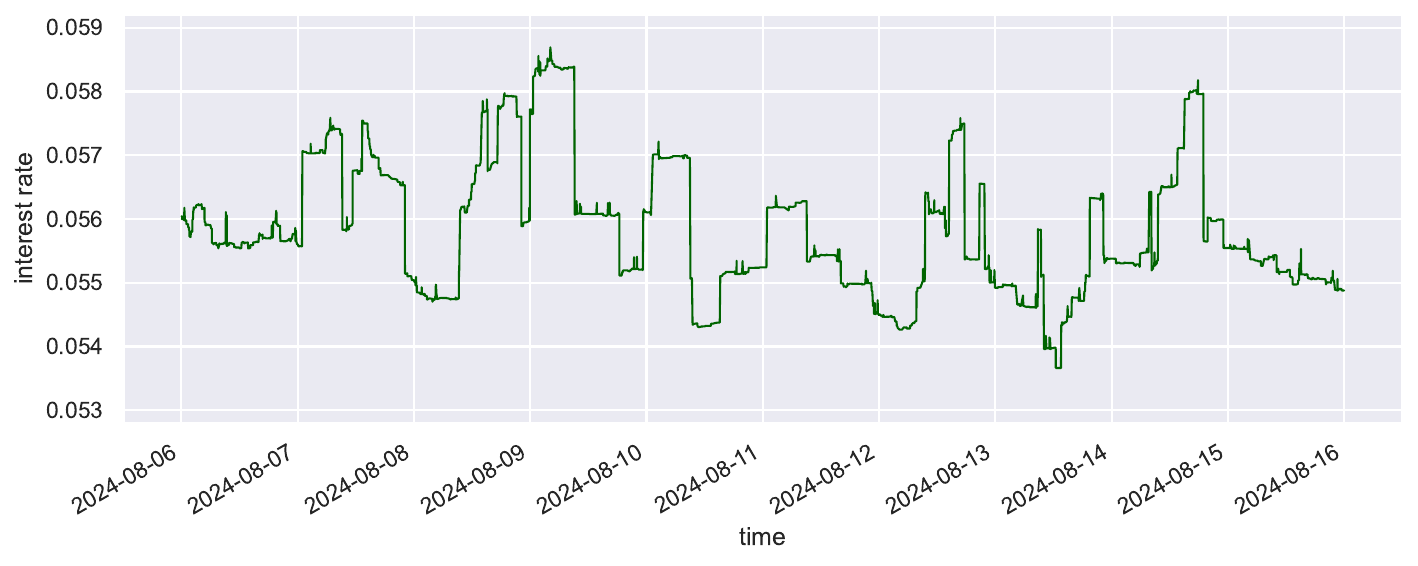}
        \caption{Interest rate}
        \label{fig:rate}
    \end{subfigure}
    \caption{Evolution of the reserves (in billion dollars), utilization and interest rate of the USDT liquidity pool from Aave v3 on the Ethereum blockchain from August 6, 2024 to August 16, 2024.}
    \label{fig:evolutions}
\end{figure}
The data used in this section is sourced from Aave v3 and retrieved through The Graph service.\footnote{\url{https://github.com/aave/protocol-subgraphs}} This analysis examines the USDT liquidity pool on the Ethereum blockchain, which is one of the largest and most active pools across the versions of Aave and the blockchains it operates on. The data characterizing the pool (e.g., the amount of liquidity supplied and borrowed) is updated and collected each time a user interacts with the smart contract (e.g., by depositing or borrowing crypto-assets). This analysis is conducted over a ten-day period, spanning from August 6, 2024 to August 16, 2024 which corresponds to almost $72000$ blocks. Market data from lending protocols are subject to frequent regime changes, driven by evolving market conditions, external events and protocol governance. Since our model is designed to capture local dynamics, we restrict the analysis to a short time window to ensure consistency with this assumption. The evolution of reserves, utilization and interest rate of the USDT liquidity pool during this period is presented in Figures \ref{fig:reserves}, \ref{fig:utilization} and \ref{fig:rate}, respectively.

For each block $i$, the variation in utilization generated by the next block is calculated as $\Delta U^{i} = U^{i+1} - U^{i}$. This total variation in utilization is then divided into increments of size $\delta$ and the number of increments, denoted by $n^{i}$, is determined using the floor and ceil functions:
\begin{equation}
n^{i} = \left\{
    \begin{array}{ll}
        \lfloor \frac{\Delta U^{i}}{\delta} \rfloor & \mbox{if } \Delta U^{i} \geq 0 \\
        \lceil \frac{\Delta U^{i}}{\delta} \rceil & \mbox{if } \Delta U^{i} < 0.
    \end{array}
\right.
\end{equation}
The interest rate corresponding to block $i$ is denoted by $r^{i}$ and the resulting dataset is $\big ( n^{i}, r^{i} \big )$.

Next, we discretize the interest rate domain by introducing the bins $B_{k} = [ b_{k}, b_{k+1} ), \quad k=0, \ldots, K-1$ with the final bin $B_{K} = [ b_{K-1}, b_{K} ]$. From the dynamics of the utilization rate \eqref{eq:utilization_dynamics}, the utilization rate increment is distributed according to a Skellam (or Poisson difference) distribution for a given bin. Thus, for each bin $k$, we estimate the empirical arrival intensities $\hat{\lambda}^{\pm}_{k}$ using maximum likelihood estimation. Details on the Skellam distribution, parameter estimation and confidence interval calculation are provided in Appendix \ref{appendix:Skellam}. Additionally, we define $\hat{r}_{k}$ as the empirical average interest rate for bin $k$. We select $K = 4$ bins, uniformly partitioned between the minimum and maximum interest rates of the dataset ($b_{0} = \min_{i} r^{i}$ and $b_{K} = \max_{i} r^{i}$).
\begin{figure}
    \centering
    \includegraphics[scale = 0.6]{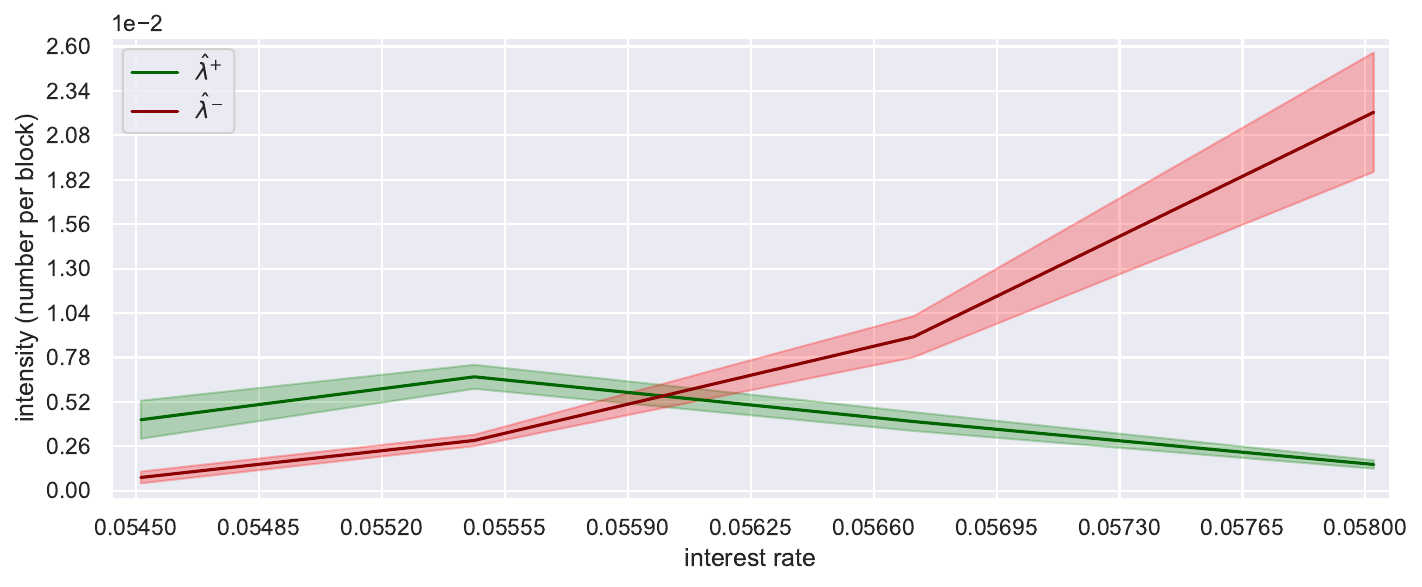}
    \caption{Calibrated market intensities with respect to the interest rate of the USDT liquidity pool from Aave v3 on the Ethereum blockchain, over the period from August 6, 2024 to August 16, 2024; $K = 4$ and $\delta = 10$ basis points.}
    \label{fig:market_intensities}
\end{figure}
Figure \ref{fig:market_intensities} illustrates the calibrated market intensities as functions of the interest rate, representing the arrival rates of events that either increase ($\hat{\lambda}^{+}$ in green) or decrease ($\hat{\lambda}^{-}$ in red) the utilization rate by $\delta = 10$ basis points. The market intensity function $\hat{\lambda}^{-}$ increases with the interest rate, reflecting the tendency of borrowers to accelerate loan repayments to avoid increased interest costs. From the perspective of liquidity providers, higher interest rates are a strong incentive to boost deposits into the liquidity pool. The market intensity function $\hat{\lambda}^{+}$ generally declines as the interest rate increases, a behavior consistent with expected market dynamics, where rising borrowing costs curtail the demand for new loans. The intersection of these two curves at $0.056$ represents an equilibrium, where the forces driving increases and decreases in utilization are balanced. At this rate, participants are equally likely to borrow or repay.

We set boundary interest rates at $r_{\text{min}} = 0$ and $r_{\text{max}} = 0.25$. Linear interpolation of market intensities within this range produces negative values and is therefore not applied. Instead, market intensities are piecewise-linearly interpolated and linearly extrapolated. To ensure non-negative values, boundary intensities are floored at zero. Additionally, $\hat{\lambda}^{-}$ is enforced to be non-increasing. The post-processed market intensities, including boundary values, are given in Table \ref{tab:market_data}.
\begin{table}
\centering
\begin{tabular}{ ccccccc }
\toprule
$\hat{r}$ & $0$ & $0.0545$ & $0.0555$ & $0.0567$ & $0.058$ & $0.25$ \\
\midrule
$\hat{\lambda}^{+}$ & $0.0067$ & $0.0067$ & $0.0067$ & $0.0041$ & $0.0015$ & $0$ \\
$\hat{\lambda}^{-}$ & $0$ & $0.0008$ & $0.0029$ & $0.009$ & $0.0222$ & $1.9485$ \\
\bottomrule
\end{tabular}
\caption{Post-processed market intensities with respect to the interest rate of the USDT liquidity pool from Aave v3 on the Ethereum blockchain, over the period from August 6, 2024 to August 16, 2024; $K = 4$ and $\delta = 10$ basis points.}
\label{tab:market_data}
\end{table}

\subsection{Architecture and detailed training procedure}\label{section:architecture}

We implement the method described in Section \ref{section:non_linear_intensities} with TensorFlow v2. The neural network employed consists of six layers: one input layer ($2$-dimensional), four hidden layers ($64$-dimensional) and one output layer ($1$-dimensional). The activation function used in the input and hidden layers is the Parametric Rectified Linear Unit (PReLU), while a linear activation function is applied to the output layer. Instead of considering a single neural network that takes both the utilization rate and time as inputs, an alternative approach is to use a family of neural networks where each sub-network is assigned a time step. The methodology employing a single neural network is selected because it produces more stable results. Similar conclusions have been drawn in other frameworks such as \citet{germain2021neural} and \citet{fecamp2019risk}.

Inspired by the algorithm proposed in \citet{cuchiero2020generative}, the training process follows a two-phase procedure. The first phase is a rapid learning phase, where a high learning rate of $10^{-3}$ is used alongside a small number of trajectories $(2 500)$. This is followed by a refinement phase with a lower learning rate of $10^{-4}$ and a larger number of trajectories $(25 000)$. In the second phase, a stopping criterion is considered based on a validation sample consisting of $250 000$ trajectories. Every $10$ iterations, the total loss function is computed using the validation sample. If the difference in loss is smaller than a predefined threshold of $10^{-7}$, the training process is terminated. In both phases, we use $10$ epochs, as recommended by \citet{bachouch2022deep}. The neural network is trained using the Adam optimizer \citep{kingma2014adam}. The starting utilization rate $u_{0}$ is uniformly randomized over the set $\{ \delta, \ldots, 1 - \delta \}$ during both learning phases. Regarding the smoothing parameter of the Heaviside function, $\varepsilon = 0.25$ is considered. Appendix \ref{appendix:algorithms} provides a pseudocode of the training procedure. Notably, Algorithm \ref{alg:training_procedure} outlines the overall training process, while Algorithm \ref{alg:one_step_training_procedure} details the Monte-Carlo implementation and the computation of the total loss function.

For the parametric models in Section \ref{section:parametric_rates}, we use a simplified version of the neural network training procedure. Specifically, in Algorithm \ref{alg:training_procedure}, the first training phase is omitted because of the limited number of parameters to calibrate, so only the second phase is considered. Regarding Algorithm \ref{alg:one_step_training_procedure}, while the Monte-Carlo framework remains unchanged, the convexity-preserving penalty function is excluded (i.e., $P_{T,N} = 0$). Instead, the model parameters are restricted to be non-negative using the TensorFlow \textit{NonNeg} constraint.

Numerical experiments are conducted on a node equipped with $2$ Intel\textsuperscript{\textregistered} Xeon\textsuperscript{\textregistered} Gold $6230$ Processors, $768$GB of RAM and $4$ GPU Nvidia\textsuperscript{\textregistered} Tesla\textsuperscript{\textregistered} V$100$ $32$GB of RAM.

\subsection{Interest rate models}\label{section:optimal_interest_rate_models}

\begin{figure}
    \centering
    \begin{subfigure}{\linewidth} 
        \centering
        \includegraphics[scale = 0.6]{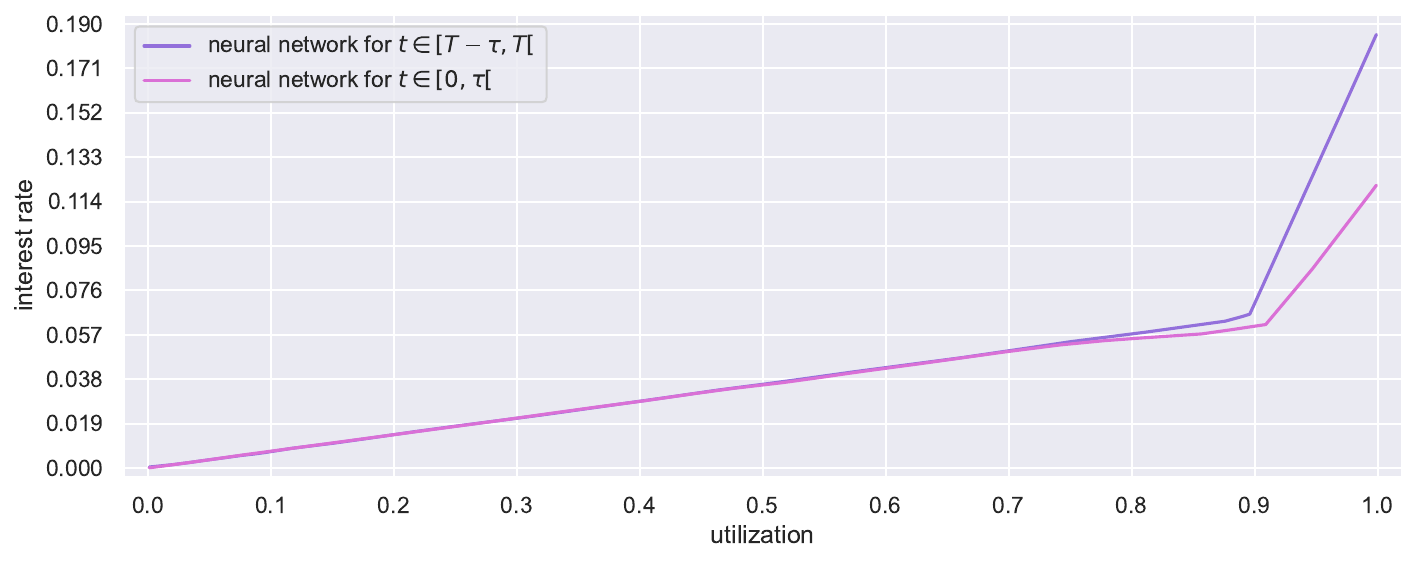}
        \caption{Optimal interest rate model}
        \label{fig:optimal_rate}
    \end{subfigure}
    \vfill
    \begin{subfigure}{\linewidth}
        \centering
        \includegraphics[scale = 0.6]{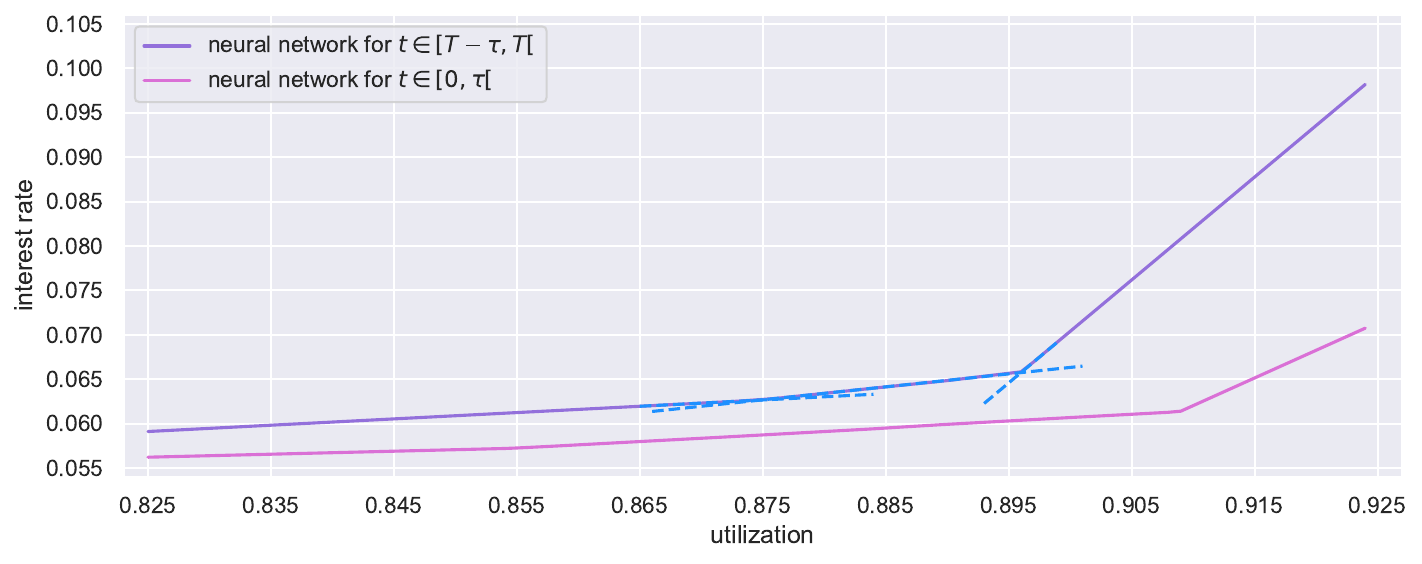}
        \caption{Zoom between $u = 0.825$ and $u = 0.925$}
        \label{fig:zoom}
    \end{subfigure}
    \caption{Interest rate as a function of utilization for $t \in [T - \tau, T[$ and $t \in [ 0, \tau [$, obtained using the neural network-based methodology; $T = 100$ blocks, $N = 100$, $J = 10$, $\delta = 0.001$, $u^{*} = 0.9$, $\phi = 7$, $\eta = 1500$ and $\bar{r} = 0$.}
    \label{fig:optimal_rates}
\end{figure}
From the intensity functions calibrated in Section \ref{section:data_processing}, we approximate the optimal interest rate model using the neural network-based methodology presented in Section \ref{section:non_linear_intensities}. The neural network learns the interest rate for $t \in [ 0, T [$ owing to the Euler scheme employed. Therefore, we present it with respect to the utilization rate at the initial period ($ t \in [ 0, \tau [ $) and the terminal one ($t \in [T - \tau, T[$), as shown in Figure \ref{fig:optimal_rate}. The resulting interest rate curves are piecewise-linear functions of utilization. More precisely, the curves are locally piecewise-linear around the target utilization, as emphasized in Figure \ref{fig:zoom}. As a result, the piecewise-linear nature of the intensity functions is directly reflected in the neural network-based interest rate model.

\begin{figure}
    \centering
    \includegraphics[scale = 0.6]{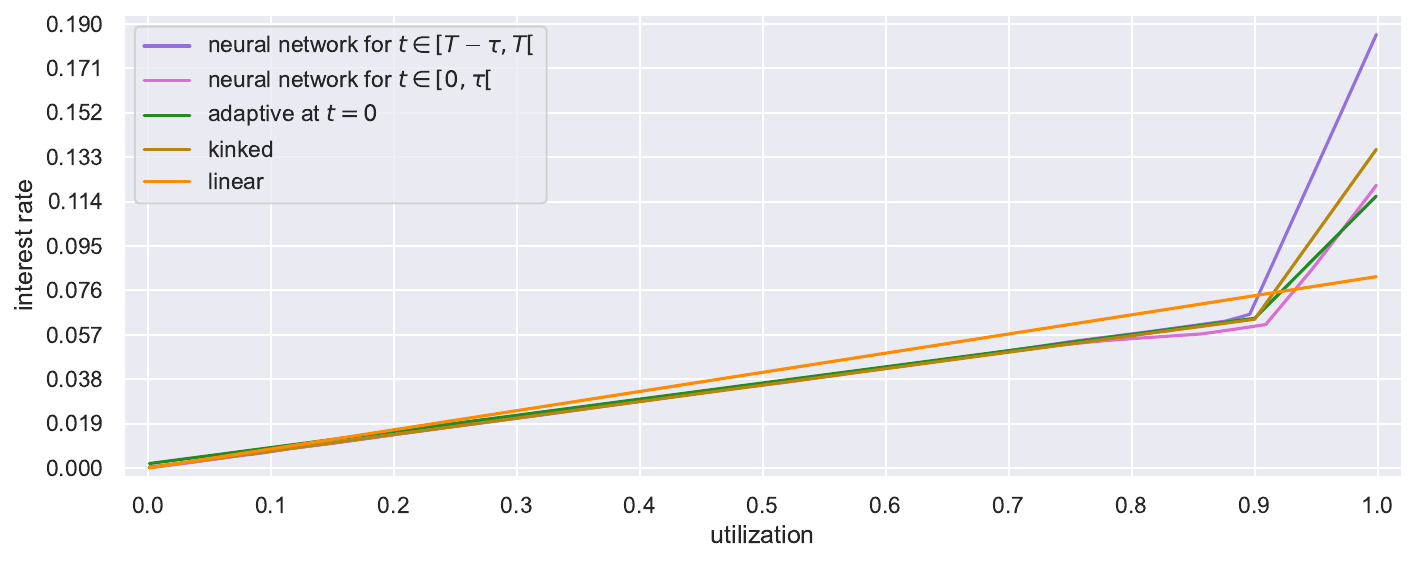}
    \caption{Linear, kinked, adaptive at $t = 0$ and neural network-based for $t \in [T - \tau, T[$ and $t \in [ 0, \tau [$ models as a function of utilization; $T = 100$ blocks, $N = 100$, $J = 10$, $\delta = 0.001$, $u^{*} = 0.9$, $\phi = 7$, $\eta = 1500$ and $\bar{r} = 0$.}
    \label{fig:all_rates}
\end{figure}
Next, after calibrating the parametric models, we compare the linear, kinked and adaptive models to the neural network-based one. The models obtained are illustrated in Figure \ref{fig:all_rates} with the adaptive model presented at initial time ($t = 0$). Below the target utilization rate, the kinked model closely aligns with the terminal interest rate curve of the neural network-based model. Additionally, the shape of the kinked model beyond the target level falls between the initial and terminal rate curves. In contrast, the linear model generally produces higher interest rates across most utilization levels compared to both curves. However, as utilization approaches $100 \%$, the linear model does not increase as sharply as the neural network-based model, potentially underpricing the risk associated with near-full utilization. A key challenge of the linear model is balancing the dual objectives of fostering liquidity pool activity and effectively managing liquidity risk with only one degree of freedom. The calibrated parameters of the linear and kinked models are presented in Table \ref{tab:linear_kinked_params}.
\begin{table}
\centering
\begin{tabular}{ cccc }
\toprule
parameter & $r_{base}$ & $r_{slope1}$ & $r_{slope2}$ \\
\midrule
linear & $0$ & $0.0738$ & $\cdot$ \\
kinked & $0$ & $0.0634$ & $0.0734$ \\
\bottomrule
\end{tabular}
\caption{Calibrated parameters of the linear and kinked models; $T = 100$ blocks, $N = 100$, $J = 10$, $\delta = 0.001$, $u^{*} = 0.9$, $\phi = 7$, $\eta = 1500$ and $\bar{r} = 0$.}
\label{tab:linear_kinked_params}
\end{table}
The adaptive model, like the kinked one, closely aligns with the terminal interest rate curve of the neural network-based model below the target utilization rate.
Beyond this level, the adaptive interest rate curve exhibits a lower slope than the kinked and neural network-based curves. However, this is counterbalanced by the dynamic nature of the adaptive model. The calibrated parameters of the adaptive rate are presented in Table \ref{tab:adaptive_params}.
\begin{table}
\centering
\begin{tabular}{ ccccc }
\toprule
parameter & $r^{\text{target}}_{0}$ & $k_{p}$ & $k_{d_{1}}$ & $k_{d_{2}}$ \\
\midrule
adaptive & $0.0641$ & $0.0015$ & $0.0294$ & $1.825$ \\
\bottomrule
\end{tabular}
\caption{Calibrated parameters of the adaptive model; $T = 100$ blocks, $N = 100$, $J = 10$, $\delta = 0.001$, $u^{*} = 0.9$, $\phi = 7$, $\eta = 1500$ and $\bar{r} = 0$.}
\label{tab:adaptive_params}
\end{table}

\subsection{Risk-adjusted PnL analysis}\label{section:penalized_pnl_analysis}

\begin{figure}
    \centering
    \begin{subfigure}{\linewidth} 
        \centering
        \includegraphics[scale = 0.6]{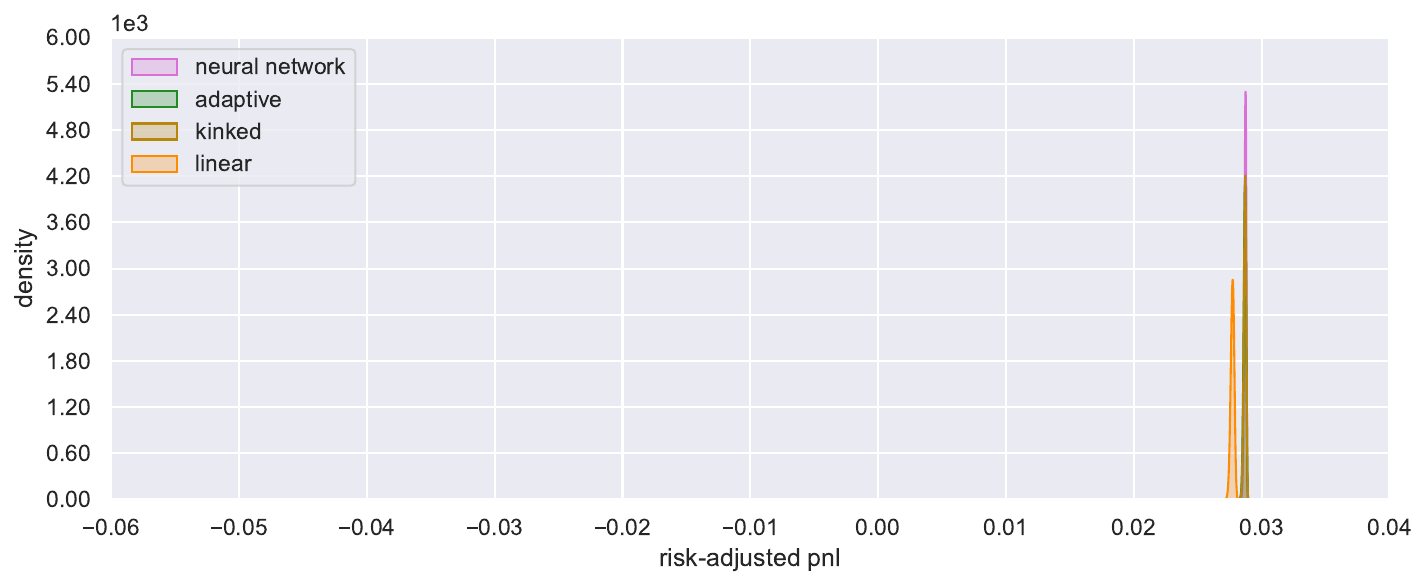}
        \caption{$u_{0} = 0.9$}
        \label{fig:penalized_pnl_90}
    \end{subfigure}
    \vfill
    \begin{subfigure}{\linewidth}
        \centering
        \includegraphics[scale = 0.6]{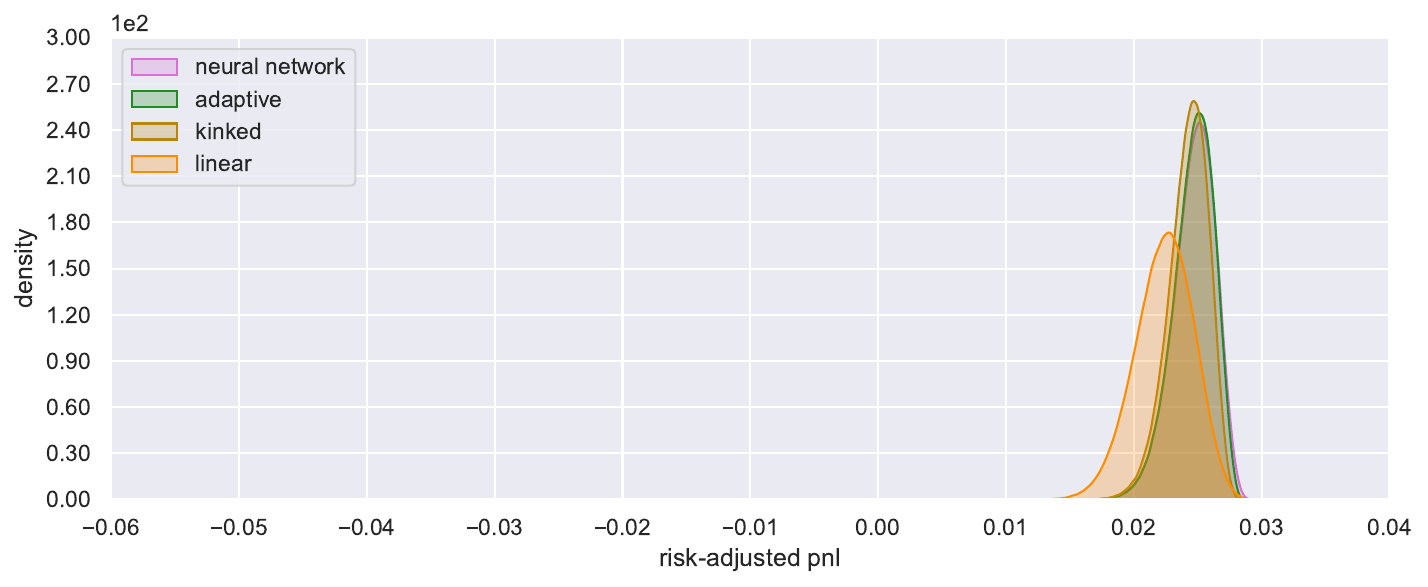}
        \caption{$u_{0} = 0.95$}
        \label{fig:penalized_pnl_95}
    \end{subfigure}
    \vfill
    \begin{subfigure}{\linewidth}
        \centering
        \includegraphics[scale = 0.6]{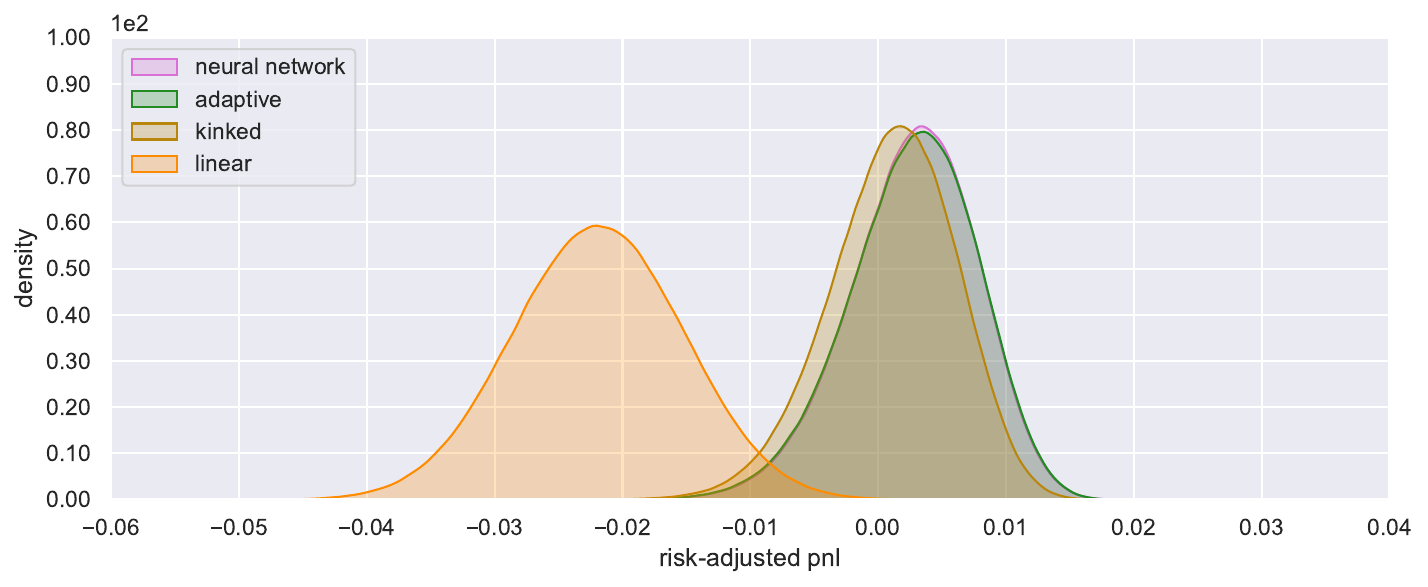}
        \caption{$u_{0} = 1$}
        \label{fig:penalized_pnl_100}
    \end{subfigure}
    \caption{Densities of the risk-adjusted PnLs for the linear, kinked, adaptive and neural network-based models; $T = 100$ blocks, $N = 100$, $J = 10$, $\delta = 0.001$, $u^{*} = 0.9$, $\phi = 7$, $\eta = 1500$ and $\bar{r} = 0$. A total of $10^{6}$ numerical simulations are performed with initial utilization rate set at $u_{0} = 0.9$, $0.95$ and $1$.}    
    \label{fig:pnls_penalized}
\end{figure}
After presenting the calibrated models in Section \ref{section:optimal_interest_rate_models}, we evaluate and compare the risk-adjusted PnL of the lenders, as defined by the loss function \eqref{eq:loss_function} given the interest rate model. A total of $10^{6}$ numerical simulations are performed, for initial utilization rate set at $u_{0} = 0.9$, $0.95$ and $1$. To compare the risk-adjusted PnLs, we compute the average, standard deviation and percentiles, expressed in basis points and presented in Table \ref{tab:pnls_penalized}. Figures \ref{fig:penalized_pnl_90}, \ref{fig:penalized_pnl_95} and \ref{fig:penalized_pnl_100} show the densities of the risk-adjusted PnLs given the initial utilization rates.
\begin{table}
\centering
\begin{tabular}{ cccccccccc }
\toprule
& \multicolumn{3}{c}{average} & \multicolumn{3}{c}{5\%-percentile} & \multicolumn{3}{c}{95\%-percentile} \\
\cmidrule(r){2-4} \cmidrule(r){5-7} \cmidrule(r){8-10}
$u_{0}$ & $0.9$ & $0.95$ & $1$ & $0.9$ & $0.95$ & $1$ & $0.9$ & $0.95$ & $1$ \\
\midrule
neural network &
287 (1) & 247 (17) & 27 (50) &
286 & 217 & -60 &
288 & 271 & 104 \\
adaptive &
287 (1) & 246 (16) & 27 (50) &
285 & 217 & -61 &
288 & 270 & 104 \\
kinked &
287 (1) & 242 (16) & 10 (50) &
285 & 214 & -76 &
288 & 266 & 87 \\
linear &
277 (1) & 223 (23) & -219 (67) &
275 & 184 & -329 &
279 & 259 & -110 \\
\bottomrule
\end{tabular}
\caption{Statistical metrics of the risk-adjusted PnLs (in basis points) for the linear, kinked, adaptive and neural network-based models; $T = 100$ blocks, $N = 100$, $J = 10$, $\delta = 0.001$, $u^{*} = 0.9$, $\phi = 7$, $\eta = 1500$ and $\bar{r} = 0$. The standard deviation is given in parentheses. A total of $10^{6}$ numerical simulations are performed with initial utilization rate set at $u_{0} = 0.9$, $0.95$ and $1$.}
\label{tab:pnls_penalized}
\end{table}
The linear model significantly underperforms relative to the other models across all initial utilization rates. When the initial utilization rate is fixed at $u_{0} = 0.9$, the risk-adjusted PnLs of the neural network-based, adaptive and kinked models are almost identical, with at most a one basis point difference. By increasing the initial utilization rate to $u_{0} = 0.95$, the neural network-based model outperforms the others, closely followed by the adaptive model, with the risk-adjusted PnL difference compared to the neural network-based model being one basis point on average. The performance of the kinked model deteriorates relative to the neural network-based and adaptive models. For an initial utilization rate of $u_{0} = 1$, the neural network-based and adaptive models exhibit similar performance, while the kinked model underperforms significantly, with an average risk-adjusted PnL reduced by more than half compared to the neural network-based and adaptive models.

We also study the risk-adjusted PnLs for an initial utilization rate $u_{0}$ uniformly randomized over the set $\{ \delta, \ldots, 1 - \delta \}$ as in the training phase \ref{section:architecture}. The average, standard deviation and percentiles are presented in Table \ref{tab:pnl_penalized_randomized}. For a randomized initial utilization rate, the risk-adjusted PnLs of the neural network-based, kinked and adaptive models are similar, largely due to the trajectories where $u_{0} \leq u^{*}$. In these cases, the interest rate models behave similarly, as illustrated in Figure \ref{fig:all_rates}, resulting in nearly identical overall performance. The linear model remains the underperformer. In Appendix \ref{appendix:pnl_analysis}, we provide an analogous analysis of the PnL, which represents the wealth accumulated by the lenders during the period, excluding the penalties.
\begin{table}
\centering
\begin{tabular}{ ccccc }
\toprule
& average &  $5 \%$-percentile & $95 \%$-percentile \\
\midrule
neural network & 108 (93) & 1 & 278 \\
adaptive & 108 (93) & 1 & 278 \\
kinked & 108 (93) & 1 & 277 \\
linear & 100 (94) & 0 & 270 \\
\bottomrule
\end{tabular}
\caption{Statistical metrics of the risk-adjusted PnLs (in basis points) for the linear, kinked, adaptive and neural network-based models; $T = 100$ blocks, $N = 100$, $J = 10$, $\delta = 0.001$, $u^{*} = 0.9$, $\phi = 7$, $\eta = 1500$ and $\bar{r} = 0$. The standard deviation is given in parentheses. A total of $10^{6}$ numerical simulations are performed with an initial utilization rate $u_{0}$ uniformly randomized over the set $\{ \delta, \ldots, 1 - \delta \}$.}
\label{tab:pnl_penalized_randomized}
\end{table}

\section{Discussion and conclusion}

In the spirit of \citet{bertucci2024agents}, we introduce a model for a liquidity pool by specifying the dynamics of the utilization rate using point processes with interest rate-dependent intensities. We then propose an interest rate model as the optimal solution to a stochastic control problem that maximizes expected lender wealth over a finite horizon, while incorporating risk penalties for liquidity risk and interest rate stabilization.

First, we derive a closed-form optimal interest rate model by assuming that intensities are linear functions of the interest rate. Under this assumption and for a very short maturity, the optimal interest rate model is a piecewise-linear function of the utilization rate with two distinct slopes which corresponds to the kinked model, introduced in \citet{whitepaper2020aavev1}. Next, we relax the linear assumption and use deep learning techniques to approximate the optimal interest rate model. We also propose a method to estimate the intensity functions based on the Skellam distribution. The methodology is illustrated using block-by-block historical data from the USDT liquidity pool of the Aave v3 protocol on the Ethereum blockchain over a short time window. By considering piecewise-linear intensities, the neural network-based interest rate model is also piecewise-linear, demonstrating its ability to adapt to diverse market conditions.

We acknowledge that implementing such interest rate models on a blockchain is challenging, notably because of the computational cost. Given this limitation, we also introduce a calibration procedure for estimating the optimal parameters of parametric interest rate models, using the same objective function of the stochastic control problem. This procedure is particularly relevant for the industry as it provides a robust and data-driven framework for parameter estimation. In this study, we consider the linear, kinked and adaptive \citep{adaptive2023morpho} models and compare their risk-adjusted PnLs to the deep learning-based one. In the considered market configuration, the performance of the adaptive rate model closely aligns with the deep learning-based model. This highlights the effectiveness of the adaptive model in balancing the dual objectives of fostering liquidity pool activity and effectively managing risks. Conversely, the kinked model lacks effective risk management, especially at near-full utilization. The linear model consistently underperforms the other models.

Modeling the dynamics of the utilization rate aligns with market standards, as major lending protocols rely on it to determine the interest rate. However, this approach also leads to a loss of information, as an increase in utilization may stem from either a new loan or a liquidity withdrawal, while a decrease may arise from a loan repayment or a liquidity deposit. For the sake of completeness, we should model the dynamics of the total values supplied and borrowed. This more refined modeling approach would enable the integration of collateral management and liquidation mechanisms into our framework, allowing, for instance, for the study of cascading liquidation phenomena.

In our modeling, the reference rate is fixed for both the intensity functions and the penalty in the control problem. While a reference rate could be introduced in the spirit of \citet{bastankhah2024thinking} or \citet{rivera2023equilibrium}, there is no consensus regarding the determination of this reference rate. We leave this topic for future research.

\section*{Acknowledgments}

The authors would like to thank Mohamed Frihat for fruitful discussions.

This work used HPC resources from the ``Mésocentre'' computing center of CentraleSupélec and École Normale Supérieure Paris-Saclay supported by CNRS and Région Île-de-France.

\section*{Disclosure of interest}

The authors have no competing interests to declare.

\section*{Declaration of funding}

This publication stems from a partnership between CentraleSupélec and PyratzLabs under the CIFRE (Conventions Industrielles de Formation par la Recherche) n\textsuperscript{o} 2023/1016.

\bibliography{bibliographie.bib}

\section*{Appendix}

\appendix
\addtocontents{toc}{\protect\setcounter{tocdepth}{-5}}
\renewcommand*{\thesubsection}{\Alph{subsection}}

\subsection{Consistency checks}\label{appendix:consistency_checks}

In this section, we present consistency checks conducted using synthetic data. We use the parameters and linear intensity functions from Section \ref{section:toy_example} to compare the solution obtained from the neural network-based methodology, as detailed in Section \ref{section:non_linear_intensities}, with the optimal interest rate model derived in Section \ref{section:linear_intensities}. The results demonstrate that the neural network-based methodology provides accurate estimates of the optimal interest rate model. The error metrics considered are:
\begin{equation}\label{eq:mean_error}
\text{mean error} = \frac{\delta}{(N-1) (1 - \delta)} \sum_{i=0}^{N-1} \sum_{u \in \{ \delta, \ldots, 1 - \delta \}} \big | r^{*}(u, t_{i}) - \hat{r}(u, t_{i}) \big |,
\end{equation}
and
\begin{equation}\label{eq:max_error}
\text{maximum error} = \max_{i=0, \ldots, N-1} \max_{u \in \{ \delta, \ldots, 1 - \delta \}} \big | r^{*}(u, t_{i}) - \hat{r}(u, t_{i}) \big |.
\end{equation}

\begin{figure}
    \centering
    \begin{subfigure}{\linewidth} 
        \centering
        \includegraphics[scale = 0.6]{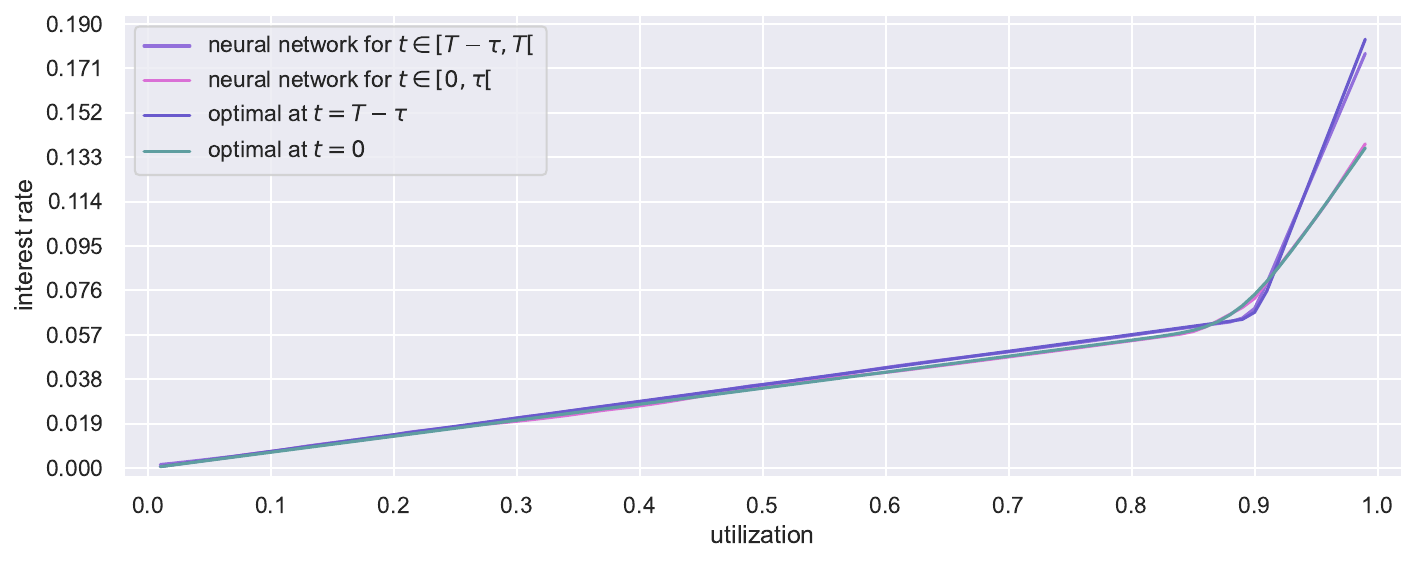}
        \caption{Interest rate models}
        \label{fig:consistency_rates}
    \end{subfigure}
    \vfill
    \begin{subfigure}{\linewidth}
        \centering
        \includegraphics[scale = 0.6]{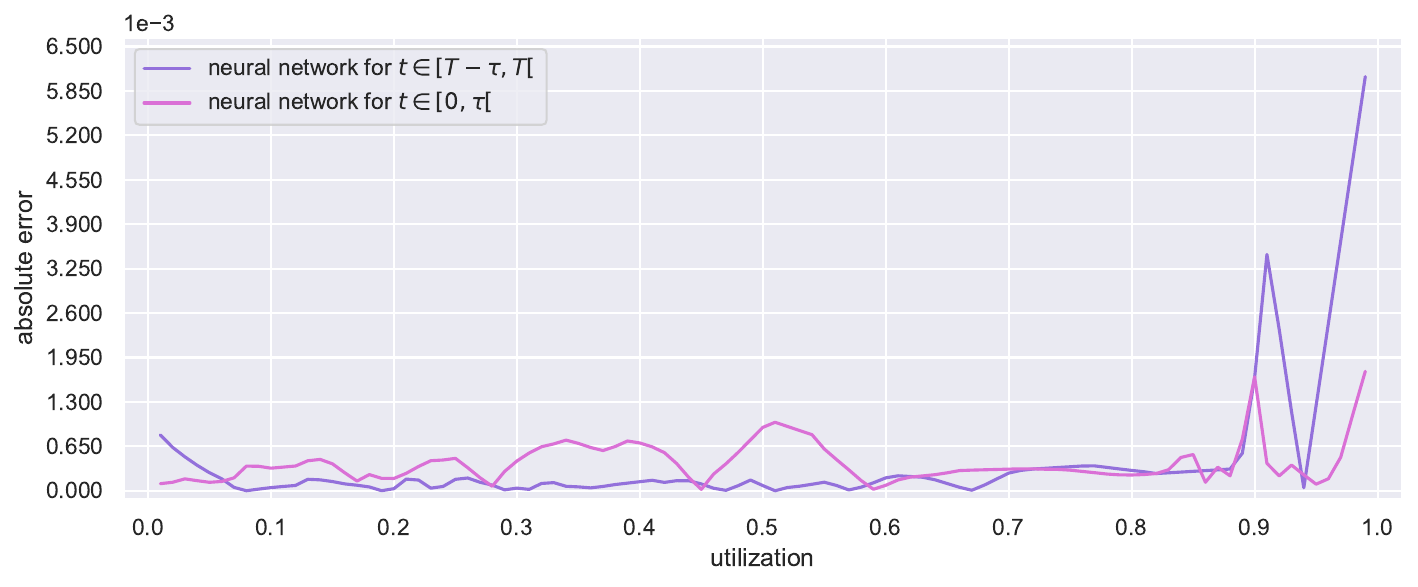}
        \caption{Absolute error}
        \label{fig:consistency_errors}
    \end{subfigure}
    \caption{Interest rate as a function of utilization for $t \in [T - \tau, T[$ and $t \in [ 0, \tau [$, obtained using the neural network-based methodology, compared to the ODEs-based model; $T = 100$ blocks, $N = 100$, $J = 10$, $\delta = 0.01$, $u^{*} = 0.9$, $\phi = 7$, $\eta = 1500$ and $\bar{r} = 0$.}
    \label{fig:rates_100_10}
\end{figure}

We vary the number of time steps $N$ and the jump trial count parameter $J$. The results, averaged over $10$ independent trainings and shown in Table \ref{tab:sanity_check_continuous_binomial}, also include the maximum probability $\bar{p}$, which represents the highest jump probability given $N$ and $J$:
\begin{equation}\label{eq:maximum_probability}
\bar{p} = \max_{r \in [r_{\text{min}}, r_{\text{max}}]} \max \big ( \lambda^{+} ( r ), \lambda^{-} ( r )\big ) \tau J^{-1}.
\end{equation}
\begin{table}
\centering
\begin{tabular}{ ccccc }
\toprule
time steps & \multicolumn{2}{c}{$N = 10$} & \multicolumn{2}{c}{$N = 100$} \\
\cmidrule(r){2-3} \cmidrule(r){4-5}
jump trial count & $J = 1$ & $J = 10$ & $J = 1$ & $J = 10$ \\
\midrule
maximum probability & $1$ & $0.1$ & $0.1$ & $0.01$ \\
\midrule
mean error (in basis points) & $4$ $(0.6)$ & $4$ $(0.8)$ & $3$ $(0.3)$ & $4$ $(0.3)$ \\
maximum error (in basis points) & $66$ $(15)$ & $69$ $(19)$ & $81$ $(14)$ & $64$ $(9)$ \\
\midrule
training time ($h:m$) & $00:12$ & $00:13$ & $01:39$ & $01:38$ \\
memory usage (in GB) & $1.9$ & $1.9$ & $3.5$ & $3.5$ \\
\bottomrule
\end{tabular}
\caption{Performance metrics for the neural network-based interest rate model relative to the ODEs-based model; $T = 100$ blocks, $\delta = 0.01$, $u^{*} = 0.9$, $\phi = 7$, $\eta = 1500$ and $\bar{r} = 0$. Performance metrics (mean error, maximum error, training time and memory usage) are averaged over $10$ independent trainings, with standard deviation in parentheses.}
\label{tab:sanity_check_continuous_binomial}
\end{table}
Table \ref{tab:sanity_check_continuous_binomial} shows that mean and maximum errors remain relatively consistent across different values of $N$ and $J$, even when the maximum probability equals $1$. Additionally, training time and memory usage increase only with the number of time steps, reflecting higher computational demands as $N$ grows. Overall, the results confirm that the neural network-based methodology accurately approximates the optimal interest rate model across the tested configurations. For a single run with $N = 100$ and $J = 10$, we present the resulting interest rates compared to the ODEs-based model and the absolute errors in Figures \ref{fig:consistency_rates} and \ref{fig:consistency_errors}, respectively.

\subsection{Algorithms}\label{appendix:algorithms}

In this section, we present the training procedure in the form of pseudocode. Algorithm \ref{alg:training_procedure} outlines the global training procedure, while Algorithm \ref{alg:one_step_training_procedure} details the Monte-Carlo implementation and the computation of the total loss function.

\begin{algorithm}
\caption{Training procedure}
\label{alg:training_procedure}
\begin{algorithmic}

\medskip

    \State 1. \textbf{Initialization} For each time step, the neural network is initialized to be a linear function of the utilization between $r_{\text{min}}$ and $r_{\text{max}}$

    \medskip

    \State 2. \textbf{First training phase}

    \medskip

    \State Set $n_{\text{max\_iter}} = 1000$ and the learning rate at $10^{-3}$

    \medskip

    \For{$k = 0, \ldots, n_{\text{max\_iter}}-1$}
        \State Generate the training sample with $n_{\text{batch}} = 2 500$

        \State Train the model using Algorithm \ref{alg:one_step_training_procedure} with the training sample, $n_{\text{epochs}} = 10$ and the learning rate
    \EndFor

    \medskip

    \State 3. \textbf{Second training phase}

    \medskip

    \State Generate the validation sample with $n_{\text{validation}} = 250 000$

    \medskip

    \State Set $continue = True$, $iter = 0$, $stopping\_threshold = 10^{-7}$, $n_{\text{min\_iter}} = 100$, $n_{\text{max\_iter}} = 1000$ and the learning rate at $10^{-4}$

    \medskip

    \While{$continue$ \textbf{and} $iter < n_{\text{max\_iter}}-1$}
        \State Generate the training sample with $n_{\text{batch}} = 25 000$

        \State Train the model using Algorithm \ref{alg:one_step_training_procedure} with the training sample, $n_{\text{epochs}} = 10$ and the learning rate

        \medskip

        \renewcommand\algorithmicthen{}
        \If{$iter \ \% \ 10 = 0$ \textbf{and} $iter \geq n_{\text{min\_iter}}$}

            \State Set $validation\_loss\_prev = validation\_loss$ and update $validation\_loss$ given the validation sample

            \medskip

            \If{$|validation\_loss - validation\_loss\_prev| < stopping\_threshold$}

                \State $continue = False$

            \EndIf

        \EndIf

        \medskip

        \State $iter = iter + 1$

        \medskip

    \EndWhile

\end{algorithmic}
\end{algorithm}

\begin{algorithm}
\caption{One-step training procedure}
\label{alg:one_step_training_procedure}
\begin{algorithmic}

\Require \\
\begin{itemize}
\renewcommand{\labelitemi}{\texttt{-}}
\item The batch size $n_{\text{batch}} \geq 1$
\item The independent realizations of the uniformly distributed random variable over the interval $(0, 1)$ $Z^{j,k,\pm}_{i} \in (0,1)^{2}$, $i \in \{ 0, \ldots, N-1 \}$, $j \in \{ 0, \ldots, J-1 \}$ and $k \in \{ 0, \ldots, n_{\text{batch}}-1 \}$
\item The starting utilization rates $u^{k}_{0} \in (0,1)$, $k \in \{ 0, \ldots, n_{\text{batch}}-1 \}$
\item The number of epochs $n_{\text{epochs}} \geq 1$
\item The learning rate
\end{itemize}

\Ensure \\
\begin{itemize}
\renewcommand{\labelitemi}{\texttt{-}}
\item Estimate of the optimal interest rate model $\hat{r}$
\item The total loss function
\end{itemize}

\medskip

\State Set $U^{k}_{0} = u^{k}_{0}$ and $X^{k}_{0} = Q^{k}_{0} = 0$, $ k \in \{ 0, \ldots, n_{\text{batch}}-1 \}$ \Comment{The trajectories are initialized.}

\medskip

\For{$i = 0, \ldots, N-1$} \Comment{Monte-Carlo simulations are performed.}
    \For{$k = 0, \ldots, n_{\text{batch}}-1$}

        \State $r^{k}_{i} = \hat{r}(U^{k}_{i}, t_{i})$

        \medskip

        \State $X^{k}_{i+1} = X^{k}_{i} + r^{k}_{i} U^{k}_{i} \tau$

        \medskip

        \State $Q^{k}_{i+1} = Q^{k}_{i} + \phi (r^{k}_{i} - \bar{r})^{2} \tau$

        \medskip

        \State $p^{k,\pm}_{i} = \lambda^{\pm} (r^{k}_{i}) \tau J^{-1}$

        \medskip

        \State $\Delta N^{k,\pm}_{i} = \sum_{j=0}^{J-1} H^{\varepsilon} \big ( L ( p^{k,\pm}_{i} ) + L ( Z^{j,k,\pm}_{i} ) \big )$

        \medskip

        \State $U^{k}_{i+1} = U^{k}_{i} + \delta \Delta N^{k,+}_{i} - \delta \Delta N^{k,-}_{i}$

    \EndFor
\EndFor

\medskip

\State Compute:

\medskip

\State $L_{T,N} = \frac{1}{n_{\text{batch}}} \sum_{k=0}^{n_{\text{batch}}-1} \big [ X^{k}_{N} - \psi(U^{k}_{N}) - Q^{k}_{N} \big ]$ \\ \\ \Comment{$L_{T,N}$ is an estimate of the objective function.}

\medskip

\State $P_{T,N} = \sum_{i=0}^{N-1} \sum_{u \in \{ u^{*}, \ldots, 1 - \delta \}} \min \big ( \hat{r}(u + \delta, t_{i}) - 2 \hat{r}(u, t_{i}) + \hat{r}(u - \delta, t_{i}), 0 \big )$ \\ \\ \Comment{$P_{T,N}$ is the convexity-preserving penalty.}

\medskip

\State Train the neural network using $n_{\text{epochs}}$ and the learning rate:

\medskip

\State $\hat{r} = \underset{r}{\mathrm{argmax}} \ \frac{L_{T,N}}{T} - \frac{P_{T,N}}{N}$ \Comment{The quantities have been normalized for the sake of consistency.}

\end{algorithmic}
\end{algorithm}

\subsection{Skellam distribution}\label{appendix:Skellam}

Skellam distribution \citep{skellam1946frequency} is the probability distribution of the difference between two independent Poisson-distributed random variables. Let $N^{+}$ and $N^{-}$ be two independent random variables such that $N^{+} \sim \text{Poisson}(\lambda^{+})$ and $N^{-} \sim \text{Poisson}(\lambda^{-})$ where $\lambda^{+}$ and $\lambda^{-}$ are the intensity rates of the respective distribution. The Skellam distribution, denoted by $\text{Skellam}(\mu^{+}, \mu^{-})$, describes the random variable $X = N^{+} - N^{-}$, which represents the difference between the two Poisson counts. The probability mass function of the Skellam distribution is given by:
\begin{equation}\label{eq:0}
\mathbb{P}(X = x) = e^{-\lambda^{+} - \lambda^{-}} (\frac{\lambda^{+}}{\lambda^{-}})^{\frac{x}{2}} I_{x} (2 \sqrt{\lambda^{+} \lambda^{-}}),
\end{equation}
where $I_{x}$ is the modified Bessel function of the first kind:
\begin{equation}\label{eq:0_bis}
I_{x} (\lambda) = \big ( \frac{\lambda}{2} \big )^{x} \sum^{+\infty}_{k=0} \frac{(\frac{\lambda^{2}}{4})^{k}}{k! (x+k)!}.
\end{equation}

Let $x_{1}, \ldots, x_{n}$ be an independent sample of size $n$ drawn from a $\text{Skellam}(\mu^{+}, \mu^{-})$ distribution. The likelihood function, denoted by $\mathcal{L}$, for the parameters $(\mu^{+}, \mu^{-})$ given the observed sample is expressed as:
\begin{equation}\label{eq:1}
\mathcal{L} = \prod^{n}_{i=1} e^{-\lambda^{+} - \lambda^{-}} (\frac{\lambda^{+}}{\lambda^{-}})^{\frac{x_{i}}{2}} I_{x_{i}} (2 \sqrt{\lambda^{+} \lambda^{-}}),
\end{equation}
and the log-likelihood is given by:
\begin{equation}\label{eq:1_bis}
\log (\mathcal{L}) = - n \lambda^{+} - n \lambda^{-} + \frac{1}{2} \log (\frac{\lambda^{+}}{\lambda^{-}}) \sum^{n}_{i=1} x_{i} + \sum^{n}_{i=1} \log (I_{x_{i}} (2 \sqrt{\lambda^{+} \lambda^{-}})).
\end{equation}

Next, the partial derivatives of the log-likelihood are:
\begin{eqnarray}
\frac{\partial \log (\mathcal{L})}{\partial \lambda^{+}} &= - n + \frac{1}{\lambda^{+}} \sum^{n}_{i=1} x_{i} + 
\sqrt{\frac{\lambda^{-}}{\lambda^{+}}} \sum^{n}_{i=1} \frac{I_{x_{i} + 1}(2 \sqrt{\lambda^{+} \lambda^{-}})}{I_{x_{i}}(2 \sqrt{\lambda^{+} \lambda^{-}})}\label{eq:2}\\
\label{eq:3}
\frac{\partial \log (\mathcal{L})}{\partial \lambda^{-}} &= - n +
\sqrt{\frac{\lambda^{+}}{\lambda^{-}}} \sum^{n}_{i=1} \frac{I_{x_{i} + 1} (2 \sqrt{\lambda^{+} \lambda^{-}})}{I_{x_{i}} (2 \sqrt{\lambda^{+} \lambda^{-}})},
\end{eqnarray}
where we used the following formula from \citet{alzaid2010poisson}:
\begin{equation}\label{eq:1_ter}
\frac{\partial I_{x} (\lambda)}{\partial \lambda} = \frac{x}{\lambda} I_{x} (\lambda) + I_{x + 1} (\lambda).
\end{equation}

The maximum likelihood estimators, denoted by $\hat{\lambda}^{+}$ and $\hat{\lambda}^{-}$, are obtained by setting equations \eqref{eq:2} and \eqref{eq:3} to zero and solving the resulting system of nonlinear equations as outlined by \citet{alzaid2010poisson}:
\begin{equation}\label{eq:5}
-n + \frac{\hat{\lambda}^{-} + \frac{1}{n} \sum^{n}_{i=1} x_{i}}{\sqrt{(\hat{\lambda}^{-} + \frac{1}{n} \sum^{n}_{i=1} x_{i}) \hat{\lambda}^{-}}} \sum^{n}_{i=1} \frac{I_{x_{i} + 1} \big (2 \sqrt{(\hat{\lambda}^{-} + \frac{1}{n} \sum^{n}_{i=1} x_{i}) \hat{\lambda}^{-}} \big )}{I_{x_{i}} \big (2 \sqrt{(\hat{\lambda}^{-} + \frac{1}{n} \sum^{n}_{i=1} x_{i}) \hat{\lambda}^{-}} \big)} = 0,
\end{equation}
\begin{equation}\label{eq:6}
\hat{\lambda}^{+} = \hat{\lambda}^{-} + \frac{1}{n} \sum^{n}_{i=1} x_{i}.
\end{equation}

Moreover, the $95 \%$ confidence intervals of the maximum likelihood estimators are given by:
\begin{equation}\label{eq:12}
\hat{\lambda}^{+} \pm 1.96 \sqrt{\frac{\mathcal{I}_{--}}{\mathcal{I}_{++} \mathcal{I}_{--} - \mathcal{I}^{2}_{+-}}}, \quad \hat{\lambda}^{-} \pm 1.96 \sqrt{\frac{\mathcal{I}_{++}}{\mathcal{I}_{++} \mathcal{I}_{--} - \mathcal{I}^{2}_{+-}}}.
\end{equation}
where $\mathcal{I}$ denotes the Fisher information matrix, which is expressed as:
\begin{equation}\label{eq:10}
\mathcal{I} = \begin{pmatrix}
\mathcal{I}_{++} & \mathcal{I}_{+-}\\
\mathcal{I}_{+-} & \mathcal{I}_{--}
\end{pmatrix},
\end{equation}
and the components of the Fisher information matrix are:
\begin{equation}\label{eq:11}
\mathcal{I}_{++} = \mathbb{E} \Big [ - \frac{\partial^{2} \log (\mathcal{L})}{\partial (\lambda^{+})^{2}} \Big ], \quad \mathcal{I}_{--} = \mathbb{E} \Big [ - \frac{\partial^{2} \log (\mathcal{L})}{\partial (\lambda^{-})^{2}} \Big ], \quad \mathcal{I}_{+-} = \mathbb{E} \Big [ - \frac{\partial^{2} \log (\mathcal{L})}{\partial \lambda^{+} \partial \lambda^{-}} \Big ].
\end{equation}

The second order derivatives of the log-likelihood are computed as follows:
\begin{equation}\label{eq:7}
\begin{split}
\frac{\partial^{2} \log (\mathcal{L})}{\partial (\lambda^{+})^{2}} = & - \frac{1}{(\lambda^{+})^{2}} \sum^{n}_{i=1} x_{i} - \frac{1}{2} \sqrt{\frac{\lambda^{-}}{\lambda^{+}}} \frac{1}{\lambda^{+}} \sum^{n}_{i=1} \frac{I_{x_{i} + 1}}{I_{x_{i}}} \\
& + \frac{\lambda^{-}}{\lambda^{+}} \sum^{n}_{i=1} \frac{1}{(I_{x_{i}})^{2}} \Big ( \frac{1}{2 \sqrt{\lambda^{+} \lambda^{-}}} I_{x_{i} + 1} I_{x_{i}} + I_{x_{i} + 2} I_{x_{i}} - (I_{x_{i} + 1})^{2} \Big ),
\end{split}
\end{equation}
\begin{equation}\label{eq:8}
\begin{split}
\frac{\partial^{2} \log (\mathcal{L})}{\partial (\lambda^{-})^{2}} = & - \frac{1}{2} \sqrt{\frac{\lambda^{+}}{\lambda^{-}}} \frac{1}{\lambda^{-}} \sum^{n}_{i=1} \frac{I_{x_{i} + 1}}{I_{x_{i}}} \\
& + \frac{\lambda^{+}}{\lambda^{-}} \sum^{n}_{i=1} \frac{1}{(I_{x_{i}})^{2}} \Big ( \frac{1}{2 \sqrt{\lambda^{+} \lambda^{-}}} I_{x_{i} + 1} I_{x_{i}} + I_{x_{i} + 2} I_{x_{i}} - (I_{x_{i} + 1})^{2} \Big ),
\end{split}
\end{equation}
\begin{equation}\label{eq:9}
\begin{split}
\frac{\partial^{2} \log (\mathcal{L})}{\partial \lambda^{+} \partial \lambda^{-}} = & \frac{1}{2 \sqrt{\lambda^{+} \lambda^{-}}} \sum^{n}_{i=1} \frac{I_{x_{i} + 1}}{I_{x_{i}}} \\
& + \sum^{n}_{i=1} \frac{1}{(I_{x_{i}})^{2}} \Big ( \frac{1}{2 \sqrt{\lambda^{+} \lambda^{-}}} I_{x_{i} + 1} I_{x_{i}} + I_{x_{i} + 2} I_{x_{i}} - (I_{x_{i} + 1})^{2} \Big ).
\end{split}
\end{equation}

\subsection{PnL analysis}\label{appendix:pnl_analysis}

\begin{figure}
    \centering
    \begin{subfigure}{\linewidth} 
        \centering
        \includegraphics[scale = 0.6]{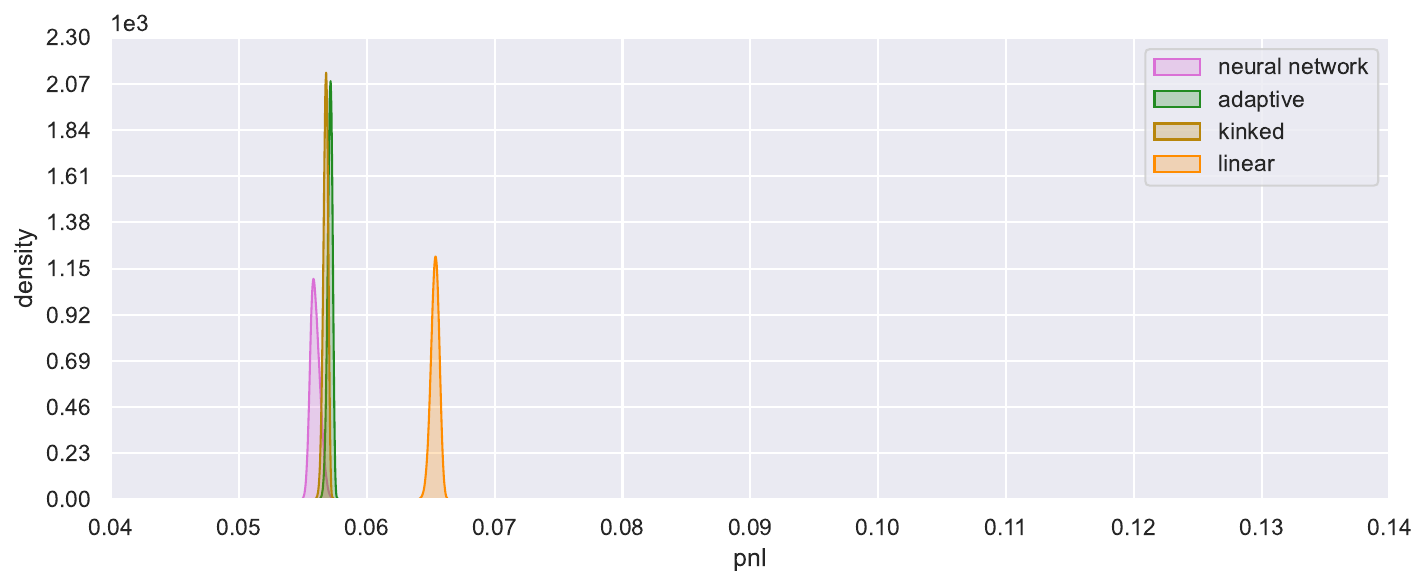}
        \caption{$u_{0} = 0.9$}
        \label{fig:pnl_90}
    \end{subfigure}
    \vfill
    \begin{subfigure}{\linewidth}
        \centering
        \includegraphics[scale = 0.6]{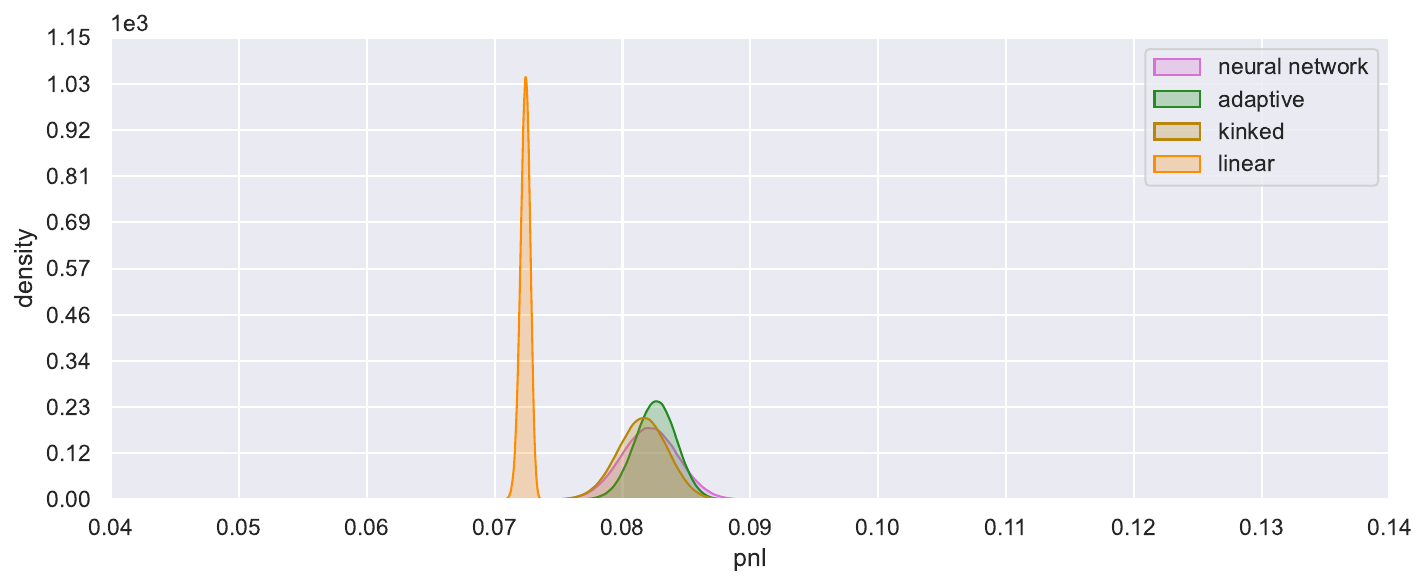}
        \caption{$u_{0} = 0.95$}
        \label{fig:pnl_95}
    \end{subfigure}
    \vfill
    \begin{subfigure}{\linewidth}
        \centering
        \includegraphics[scale = 0.6]{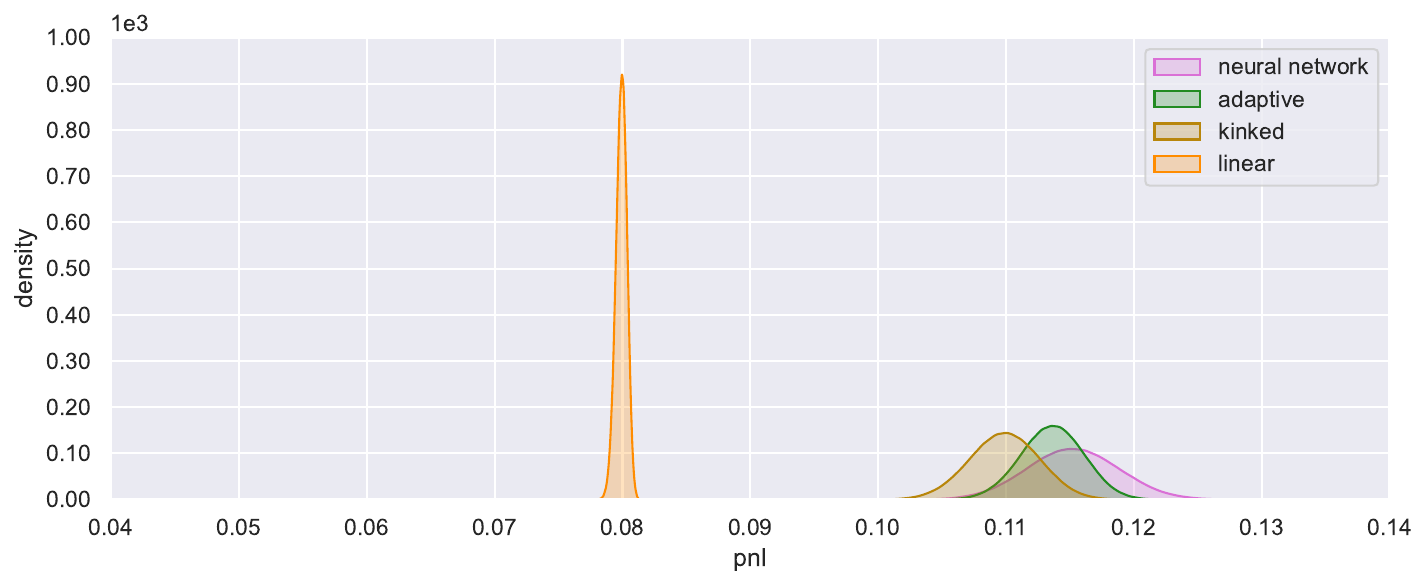}
        \caption{$u_{0} = 1$}
        \label{fig:pnl_100}
    \end{subfigure}
    \caption{Densities of the PnLs for the linear, kinked, adaptive and neural network-based models; $T = 100$ blocks, $N = 100$, $J = 10$, $\delta = 0.001$, $u^{*} = 0.9$, $\phi = 7$, $\eta = 1500$ and $\bar{r} = 0$. A total of $10^{6}$ numerical simulations are performed with initial utilization rate set at $u_{0} = 0.9$, $0.95$ and $1$.}
    \label{fig:pnls}
\end{figure}
To complete Section \ref{section:penalized_pnl_analysis}, we present a PnL analysis given the interest rate model. A total of $10^{6}$ numerical simulations are also performed, for initial utilization rate set at $u_{0} = 0.9$, $0.95$ and $1$. The densities of the PnLs, given the initial utilization rate, are presented in Figures \ref{fig:pnl_90}, \ref{fig:pnl_95} and \ref{fig:pnl_100} and the associated statistical metrics in Table \ref{tab:pnls}. We also present statistical metrics of the PnLs for an utilization rate $u_{0}$ uniformly randomized over the set $\{ \delta, \ldots, 1 - \delta \}$ in Table \ref{tab:pnl_randomized}.
\begin{table}
\centering
\begin{tabular}{ cccccccccc }
\toprule
& \multicolumn{3}{c}{average} & \multicolumn{3}{c}{5\%-percentile} & \multicolumn{3}{c}{95\%-percentile} \\
\cmidrule(r){2-4} \cmidrule(r){5-7} \cmidrule(r){8-10}
$u_{0}$ & $0.9$ & $0.95$ & $1$ & $0.9$ & $0.95$ & $1$ & $0.9$ & $0.95$ & $1$ \\
\midrule
neural network &
560 (4) & 821 (22) & 1152 (36) &
554 & 785 & 1093 &
567 & 858 & 1212 \\
adaptive &
571 (2) & 826 (16) & 1136 (25) &
568 & 799 & 1095 &
574 & 852 & 1176 \\
kinked &
568 (2) & 816 (20) & 1099 (28) &
565 & 783 & 1053 &
571 & 848 & 1144 \\
linear &
653 (3) & 724 (4) & 799 (4) &
648 & 717 & 792 &
659 & 730 & 806 \\
\bottomrule
\end{tabular}
\caption{Statistical metrics of the PnLs (in basis points) for the linear, kinked, adaptive and neural network-based models; $T = 100$ blocks, $N = 100$, $J = 10$, $\delta = 0.001$, $u^{*} = 0.9$, $\phi = 7$, $\eta = 1500$ and $\bar{r} = 0$. The standard deviation is given in parentheses. A total of $10^{6}$ numerical simulations are performed with initial utilization rate set at $u_{0} = 0.9$, $0.95$ and $1$.}
\label{tab:pnls}
\end{table}

\begin{table}
\centering
\begin{tabular}{ cccccc }
\toprule
& average &  $5 \%$-percentile & $95 \%$-percentile \\
\midrule
neural network & 254 (255) & 2 & 816 \\
adaptive & 253 (257) & 2 & 822 \\
kinked & 253 (253) & 2 & 813 \\
linear & 271 (240) & 2 & 724 \\
\bottomrule
\end{tabular}
\caption{Statistical metrics of the PnLs (in basis points) for the linear, kinked, adaptive and neural network-based models; $T = 100$ blocks, $N = 100$, $J = 10$, $\delta = 0.001$, $u^{*} = 0.9$, $\phi = 7$, $\eta = 1500$ and $\bar{r} = 0$. The standard deviation is given in parentheses. A total of $10^{6}$ numerical simulations are performed with an initial utilization rate $u_{0}$ uniformly randomized over the set $\{ \delta, \ldots, 1 - \delta \}$.}
\label{tab:pnl_randomized}
\end{table}
For a randomized initial utilization rate, the performance of the neural network-based, adaptive and kinked models are similar. In contrast, the linear model outperforms the others due to its higher interest rates for $u \leq u^{*}$, as illustrated in Figure \ref{fig:all_rates}. Specifically, for $u_{0} = 0.9$, the linear model outperforms, whereas for $u_{0} = 0.95$ and $u_{0} = 1$, it underperforms relative to the others.

\end{document}